\documentclass[smallextended]{svjour3}       % onecolumn (second format)
\usepackage{mathptmx}
\usepackage{amsmath}
\usepackage{graphicx}
\usepackage[english]{babel}

\journalname{Exp Fluids}

\begin{document}

\title{Combined stereoscopic wave mapping and particle image velocimetry}

\author{Quentin Aubourg \and Joel Sommeria \and Samuel Viboud \and Nicolas Mordant}
%\affiliation{Laboratoire des Ecoulements G\'eophysiques et Industriels, Universit\'e Grenoble Alpes \& CNRS, Grenoble, France}
\institute{Laboratoire des Ecoulements G\'eophysiques et Industriels, Universit\'e Grenoble Alpes \& CNRS, Grenoble, France}
\date{Received: date / Accepted: date}
\maketitle

\begin{abstract}
We describe a technique of  image analysis providing the time resolved shape of a water surface simultaneously with the three-component velocity field on this surface. The method relies on a combination of stereoscopic surface mapping and stereoscopic three-component Particle Imaging Velocimetry (PIV), using a seeding of the free surface by small polystyrene particles. The method is applied to an experiment of `weak turbulence' in which random gravity waves interact in a weakly non-linear regime. Time resolved fields of the water elevation are compared to the velocity fields on the water surface, providing direct access to the non-linear advective effects. The precision of the method is evaluated by different criteria: tests on synthetic images, consistency between several camera pairs, comparison with capacity probes. The typical r.m.s. error corresponds to 0.3 pixel in surface elevation and time displacement for PIV, which corresponds to about 0.3 mm or 1 \% in relative precision for our field of view 2 x 1.5 m$^2$.


\end{abstract}

\section{Introduction}

A variety of techniques provides the surface reconstruction of a solid or a fluid from a set of images. In this paper, we focus on the case of waves propagating at the surface of water. We aim at resolving both the water surface shape and the velocity field of gravity waves in a large scale experiment ($>10$~m). We wish to obtain a measurement resolved both in space and time in order to study the nonlinear dynamics of random wave fields. In the limit of small non-linearities, the Weak Turbulence Theory (\cite{nazarenko2011wave}) predicts that energy is transferred through resonant wave interactions. For stationary forced turbulence, analytic expressions of the wave spectrum can be derived: the so-called Kolmogorov-Zakharov spectrum. It shows that energy cascades down-scale following a power law spectrum $ E^{\eta}(\omega)\propto \omega^{-4}$ for the vertical elevation field $\eta$  of surface gravity waves. 

The method described here has been initiated with the aim of analyzing high order correlations in space and time, in order to test directly the resonant wave coupling which is at the core of the theory. The challenge is then the need for a good resolution both in time and space in order to identify accurately the components of resonant interactions. Furthermore a large amount of data is needed to get a good statistical convergence. These high order correlations have been recently analyzed on gravity-capillary waves ( \cite{aubourg2015,aubourg2016}) using an optical method called the Fourier transform profilometry (\cite{Takeda1983,Maurel}). The principle is to project a pattern on the surface of water and to demodulate the deformation of this pattern to recover the water surface deformation. The projection is made possible by the addition of white pigment that renders water optically diffusive and thus enables the image of the pattern to form very close to the water surface. However, in the case of the  large size experiments required for gravity wave studies, this method would require a tremendous amount of pigment. Methods based on speckle patterns (\cite{tanaka2002}) are also limited to small (millimetric) vertical displacements.

 Alternative solutions based on the refractive properties \cite{morris2005dynamic,Moisy2009} of the water are also ineffective due to the presence of relatively steep slopes that will induce caustics which breaks the regular correspondence between the surface displacement and the image. Other refraction-based methods (\cite{fouras2008measurement}, \cite{gomit2013} ) allow for steeper waves but require a laser sheet illumination of seeding particles inside the water layer, which is difficult in the large scale experiments considered here. 
 
  The most effective way is then to use a multi-view system that provides a reconstruction through stereoscopic algorithms \cite{prasad2000stereoscopic}. In the case of water, we can mention the work of \cite{Chatellier2010,chatellier2013parametric} or \cite{Douxchamps2005} that resolved the free surface in small scale experiments.  The implementation of these techniques requires the presence of a pattern on the free surface in order to identify the correspondence of points in the different views. This pattern can be generated by a seeding with floating particles (see \cite{Douxchamps2005}), preferably  fluorescent (see \cite{turney2009method}). For in-situ measurements, it is also possible to use directly the free surface roughness due to capillary waves or natural impurities (see \cite{Benetazzo2006,DeVries2011}).  Unfortunately, it seems very difficult to rely on such natural patterns in a laboratory experiment where the water is transparent. Seeding particles at the surface provides a suitable alternative way. This has been used to measure the velocity field at the surface by 2D PIV in laboratory experiments (\cite{weitbrecht2002}) and in rivers  (\cite{jodeau2008}), assuming the surface remains plane.
 
 To deal with deformed surfaces, different methods of pattern matching have been used to identify the same points in two stereoscopic views. In ref. \cite{Douxchamps2005}, individual particles were detected and their patterns were matched. In ref.  \cite{Chatellier2010,chatellier2013parametric}, Digital Image Correlation (DIC) was used to optimize a global functional form for the surface shape. This is appropriate for smooth deformations, as obtained in solid mechanics or in the viscous flows occurring  in small scale experiments. In the multi-scale field of our study, the global optimisation problem would involve many parameters. Therefore we rather proceed locally by optimizing cross-correlations in small sub-images, like in traditional  Particle Imaging Velocimetry (PIV). 
 
The main novelty of our method is to combine the stereoscopic reconstruction of the water surface and the stereoscopic PIV to map the velocity field on the rough surface deformed by gravity waves. This provides the horizontal velocity components in addition to the vertical one, which is itself obtained with a better precision than the time derivative of the surface elevation.  This therefore increases the measurement dynamics of the wave power spectrum, since the velocity spectrum decreases more slowly with the frequency (or the wave number) than  the surface elevation so that the former requires potentially a lesser dynamical range. In the case of wave investigation, the knowledge of the velocity also allows us to estimate directly the intensity of non linearities. Note that a similar combination of PIV and stereoscopic reconstruction has been applied to the measurement of 3D metal deformation (\cite{garcia2002}), but it was limited to small displacements. Our technique of combined stereoscopic surface reconstruction and 3 component PIV therefore seems to be new for large deformations in a rough gravity wave fields.

\begin{figure}
\includegraphics[width=14cm]{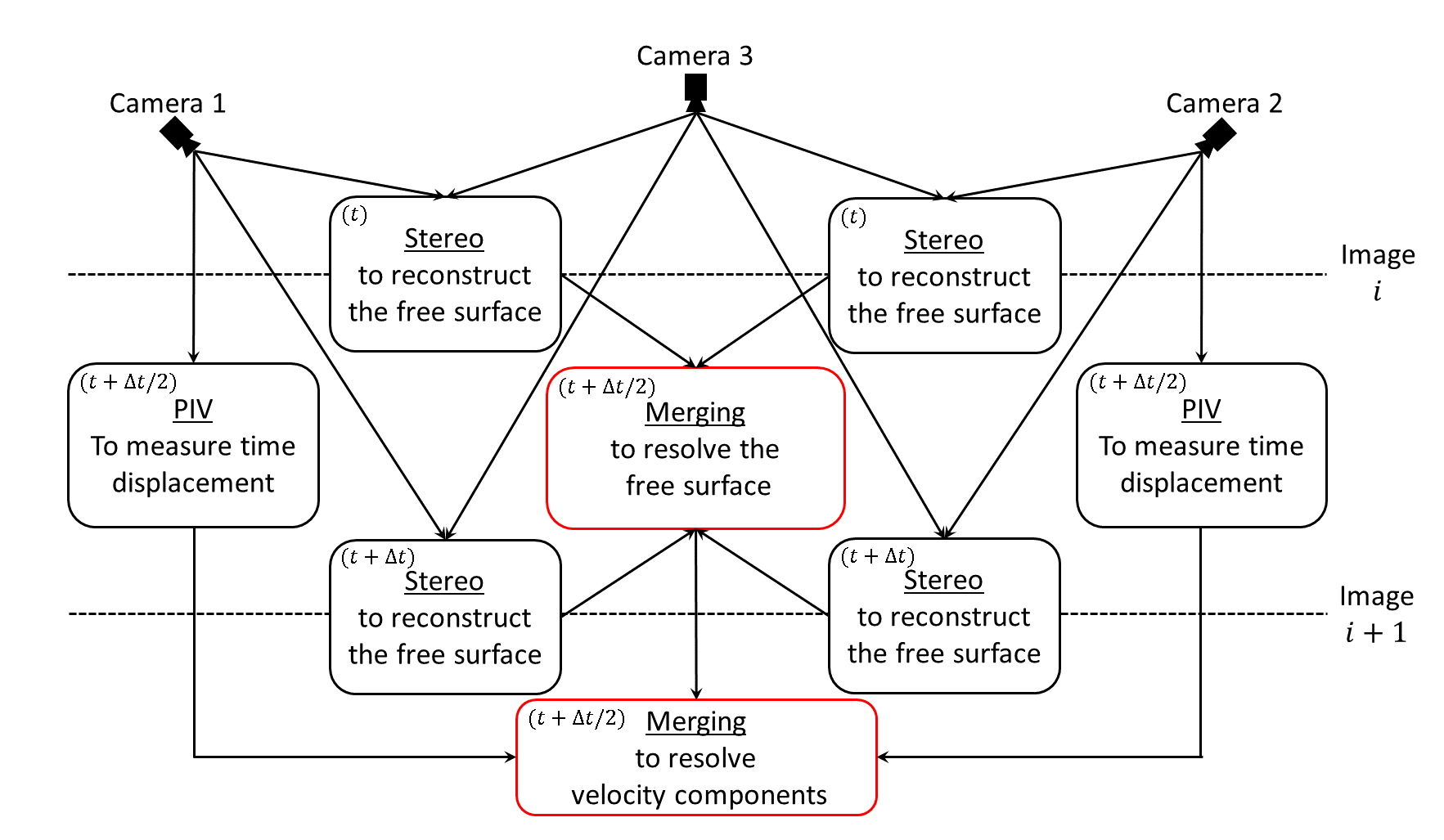}
\caption{General organization of the measurement. Three cameras are used. For the two extremes ($1$ and $2$) a PIV correlation between 2 successive times $t$ and $t+\Delta t$ is performed  to get particle displacements in sensor coordinates. At each time the two pairs (1-3) and (2-3) are used to get the free surface shape by stereoscopy. These two fields $z(x,y)$ are compared and averaged with the two fields similarly obtained at the next time step, to get the surface reconstruction at the same time $t+\Delta t/2$ as the PIV. This knowledge of  $z(x,y)$ allows us to convert the PIV displacement on the camera sensors to the 3D velocity vectors $\mathbf u(x,y,t+\Delta t/2)$ on the reconstructed surface. The whole set of operations is repeated for each frame in the time series}
\label{fig:orga}
\end{figure}

 Fig. \ref{fig:orga} shows the global procedure that we have implemented. As will be explained below, three cameras have been used to optimize the choices of viewing angles, which may be different for stereoscopic reconstruction and 3 component PIV. We use the redundancy  of different camera pairs to improve the reliability and precision, although our algorithms are limited to a single camera pair (generically labeled by the subscripts $a$ and $b$ in the following). The first step is a stereo cross-correlation between the different camera pairs in the physical space in order to reconstruct the surface $z(x,y)$.The second step consists in performing  PIV correlations on each camera in order to measure the  particle displacement in time viewed on the camera sensor. Knowing the position $z(x,y)$ from the previous step, it is then possible to merge the  PIV displacements issued from each camera  in order to construct the three component velocity field $\mathbf u (x,y,t)$ on the surface. Sections \ref{stereo} and \ref{PIVstereo} give the mathematical definition of the stereo-PIV reconstruction. Section \ref{gravitywaves} describes the experiment in the Coriolis platform where the method has been applied and shows some experimental results. The last part (\ref{Tests_precision}) investigates the accuracy of the method in the conditions of the experiment. Some keys for the practical implementation and the calibration are given in the appendix.

\section{Stereoscopic surface mapping}
%%%%%%%%%%%%%%%%%%%%
\label{stereo}

The geometric camera calibration consists in a relation between the
physical coordinates $(x,y,z)$ and the \emph{image coordinates} $(X_{a},Y_{a})$
on the camera sensor, expressed in pixel units. The latter are denoted
with the subscript $a$ anticipating that a second camera $b$ will
be used. 
\begin{equation}
X_{a}=F_{a}(x,y,z)\quad,\quad Y_{a}=G_{a}(x,y,z)\label{eq:geometric_transform}
\end{equation}
The calibration functions $F_{a}$ and $G_{a}$ depend of course on
the camera features and position, so they are also denoted with the
subscript $a$. The pinhole camera model discussed in the appendix provides
a standard choice of transform functions, but our approach is general
at this stage. 

Inverting these relations is not possible without additional information
as we have three unknown $x,y$,$z$ but only two relations. This
indetermination can be resolved if the objects of interest are contained
in a known plane, for instance the plane $z=0$, which is chosen as
the horizontal free surface at rest in our case. For each point $X_{a},Y_{a}$
on the image, we can define a line of physical points which project
on this same image point, so they cannot be distinguished with a
single camera $a$. Those satisfy the equations 
\begin{equation}
F_{a}(x,y,z)=X_{a}\equiv F_{a}(x_{a,},y_{a},0)\quad,\quad G_{a}(x,y,z)=Y_{a}\equiv G_{a}(x_{a,},y_{a},0)
\end{equation}
where we define the \emph{apparent physical coordinates $x_{a},y_{a}$}
as the intersection of this line with the reference plane\emph{.}

\begin{figure}
\includegraphics[width=10cm]{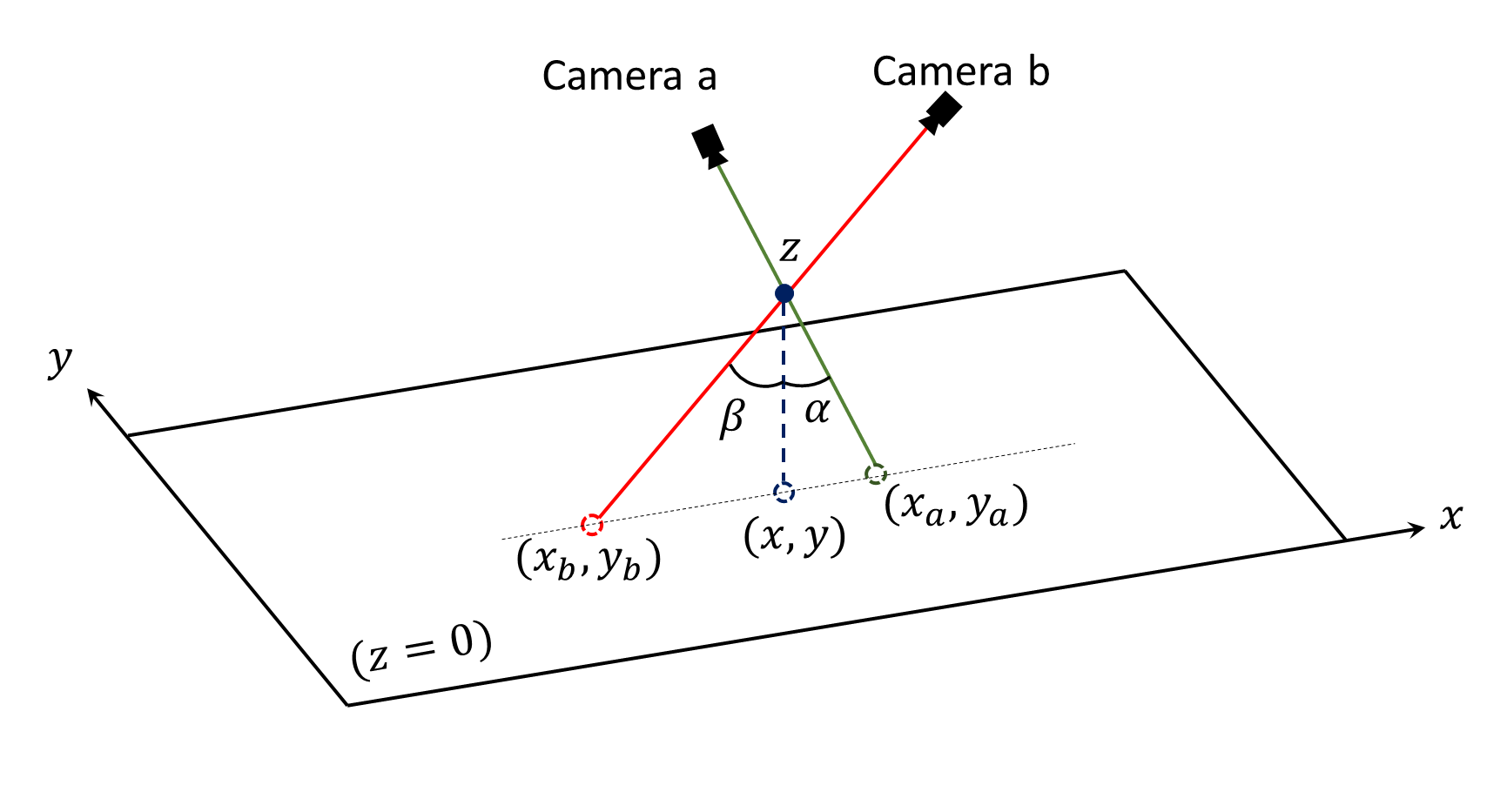}

\caption{Apparent physical coordinates $(x_a,y_a)$ and $(x_b,y_b)$  in the reference plane ($z=0$) for a point $z(x,y)$ out of the plane viewed by two camera $a$ and $b$. The angle $\alpha$ and $\beta$ are the respective incidence angles of the two cameras.}
\label{fig:reference-physical-coordinates}
\end{figure}

We can linearize these equations near the reference plane by introducing
the partial differentials of \emph{ $F_{a}$} and\emph{ $G_{a}$} at
the point $(x_{a},y_{a},0)$ , 
\begin{equation} 
\begin{array}{c}
\frac{\partial F_{a}}{\partial x}(x-x_{a})+\frac{\partial F_{a}}{\partial y}(y-y_{a})+\frac{\partial F_{a}}{\partial z}z=0\\
\frac{\partial G_{a}}{\partial x}(x-x_{a})+\frac{\partial G_{a}}{\partial y}(y-y_{a})+\frac{\partial G_{a}}{\partial z}z=0
\end{array}
\label{linearised}
\end{equation}
which defines a straight line intersecting the point \emph{$(x_{a},y_{a,},0)$},
the \emph{line of sight} of camera \emph{a} at this point (see Fig. \ref{fig:reference-physical-coordinates}). We have similar relations for camera $b$,
\begin{equation}
\begin{array}{c}
\frac{\partial F_{b}}{\partial x}(x-x_b)+\frac{\partial F_{b}}{\partial y}(y-y_{b})+\frac{\partial F_{b}}{\partial z}z=0\\
\frac{\partial G_{b}}{\partial x}(x-x_{b})+\frac{\partial G_{b}}{\partial y}(y-y_{b})+\frac{\partial G_{b}}{\partial z}z=0
\end{array}
\label{linearised_b}
\end{equation}
Because of the propagation
of light along straight lines in the air above the surface, this linear expansion is in fact exact
even away from the reference plane $z=0$. However we can use the same technique to measure internal waves at the interface between two layers of different densities. In that case optical rays are deviated by refraction and the linearized equations are only applicable as a linear approximation valid for moderate deviation from the reference. 

The two equations of Eq. \ref{linearised}  can be solved
in terms of the $z$ coordinate 
\begin{equation}
\begin{cases}
x-x_{a}=D_{xa}z\\
y-y_{a}=D_{ya}z
\end{cases}\quad\mathrm{with}\;D_{xa}=\frac{(\partial G_{a}/\partial y)(\partial F_{a}/\partial z)-(\partial F_{a}/\partial x)(\partial G_{a}/\partial z)}{(\partial G_{a}/\partial y)(\partial F_{a}/\partial x)-(\partial F_{a}/\partial y)(\partial G_{a}/\partial x)}\label{eq:line_sight}
\end{equation}
($D_{ya}$ is similarly defined by switching $x$ and $y$).  Using similarly definitions of $D_{xb}$ and $D_{yb}$ for camera $b$, we get
\begin{equation}
\begin{array}{c}
x-x_{b}=D_{xb}z\\
y-y_{b}=D_{yb}z
\end{array}
\label{eq:line_sight_b}
\end{equation}
The parameters involved can be simply interpreted as the tangent of the incidence angle for the line of sight of each camera. With the two cameras aligned
along the $x$ axis, as sketched in Fig. \ref{fig:reference-physical-coordinates}, we have indeed  $x-x_{a}=z\,\mathrm{tan}\alpha$ 
and $x-x_{b}=-z\,\mathrm{tan}\beta$, so that  $D_{xa}=\mathrm{tan}\alpha$
and $D_{xb}=-\mathrm{tan}\beta$. Along the $y$ axis, $y-y_a=y-y_b=0$ so that $D_{ya}=D_{yb}=0$. Eq. \ref{eq:line_sight} and \ref{eq:line_sight_b} provide a generalization to any orientation of the cameras, taking into account also that the viewing angle (hence the coefficients $D$) depends on the position on the image. 

For a given point with known  image coordinates, the determination of the three physical coordinates $x,y,z$  involves four equations, for instance those of Eq. \ref{linearised} and \ref{linearised_b}. This imposes a solvability constraint on the image coordinates, easily obtained by a linear combination of the four equations in Eq. \ref{eq:line_sight} and \ref{eq:line_sight_b}.
\begin{equation}
(D_{yb}\,-\,D_{ya})(x_b-x_a)\,-\,(D_{xb}-D_{xa})(y_b-y_a)\,=\,0\label{eq:solvability_cond}
\end{equation}
In the example of Fig. \ref{fig:reference-physical-coordinates}, this reduces to the coincidence $y_a=y_b$ .  However such a constraint is never exactly satisfied because of measurement errors. A classical approach is then to replace 0 by a small perturbation $\epsilon_i$  in the right hand sides of each of the equations in \ref{linearised} and \ref{linearised_b} and to minimize the sum of these perturbations squared. The minimization problem then yields a linear system of three equations with a unique solution. 

In our case however we consider that the error is solely due to the  point matching by image correlation. Assuming the set of chosen positions $(x_a,y_a)$ is exact on the first image, the error is then solely in  $(x_b,y_b)$. Imperfect geometric calibration may be another source of error, but it is smooth in $x,y$ and steady in time, so it is not critical for our application, which is to resolve the wave oscillations at rather small scale. Therefore we replace $(x_b,y_b)$  by $(x_b-\epsilon_x,y_b-\epsilon_y)$ in the previous problem, considering that $(x_b,y_b)$ are now the measured coordinates and $(\epsilon_x,\epsilon_y)$ the errors.  We assume  that $(x_a,y_a)$ are exactly known, so that the two equations in Eq. \ref{eq:line_sight} are unchanged. By contrast in the two equations of Eq. \ref{eq:line_sight_b}, we are led to add $(-\epsilon_x,-\epsilon_y)$ respectively to the right hand side of each line, which yields, by linear combinations in Eq. \ref{eq:line_sight} and \ref{eq:line_sight_b},
\begin{equation}
\begin{array}{c}
x_b-x_a=(D_{xb}-D_{xa})\,z+\epsilon_x\\
y_b-y_a=(D_{yb}-D_{ya})\,z+\epsilon_y\\
(D_{xb}-D_{xa})\,x\,=\,D_{xb}x_a\,-\,D_{xa}x_b \,+\,D_{xa} \epsilon_x\\
(D_{yb}-D_{ya})\,y\,=\,D_{yb}y_a\,-\,D_{ya}y_b\, +\,D_{ya} \epsilon_y
\end{array}
\label{eq:full system}
\end{equation}

For this system, the solvability condition Eq. \ref{eq:solvability_cond} becomes
\begin{equation}
(D_{yb}\,-\,D_{ya})\epsilon_x\,-\,(D_{xb}-D_{xa})\epsilon_y\,=(D_{yb}-D_{ya})(x_b-x_a)\,-\,(D_{xb}-D_{xa})(y_b-y_a)\ \label{eq:solvability_cond_epsilon}
\end{equation}
which defines a line in the plane $(\epsilon_x,\epsilon_y)$. We then consider the rotated error $\epsilon_X,\epsilon_Y$ projected respectively along this line and along a perpendicular direction, 
\begin{equation}
\begin{array}{c}
\epsilon_{x}=\mathrm{cos}(\phi)\,\epsilon_X\,+\,\mathrm{sin}(\phi)\,\epsilon_Y \\
\epsilon_{y}=-\mathrm{sin}(\phi)\,\epsilon_X\,+\,\mathrm{cos}(\phi)\,\epsilon_Y
\end{array}
\label{eq:epsilon_phi}
\end{equation}

\begin{equation}
\begin{array}{c}
\mathrm{cos}\phi=-\frac{D_{xb}-D_{xa}}{[(D_{xb}-D_{xa})^2+(D_{yb}-D_{ya})^2]^{1/2}}\\
\mathrm{sin}\phi=\frac{D_{yb}-D_{ya}}{[(D_{xb}-D_{xa})^2+(D_{yb}-D_{ya})^2]^{1/2}}
\end{array}
\label{eq:epsilon_phi_D}
\end{equation}

Introducing that in Eq. \ref{eq:solvability_cond_epsilon} yields
\begin{equation}
\epsilon_Y=\frac{(D_{yb}-D_{ya})\,(x_b-x_a)\,-\,(D_{xb}-D_{xa})\,(y_b-y_a)}{[(D_{xb}-D_{xa})^2+(D_{yb}-D_{ya})^2]^{1/2}}\label{eq:error_X}
\end{equation}
while the transverse error $\epsilon_X$  is undetermined. It should satisfy a probability distribution which can be assumed to be centered around 0. In the case of two lines of sight aligned with $x$, $D_{yb}-D_{ya}=0$, and  $\epsilon_Y$ reduces to the mismatch $y_b-y_a$, while $\epsilon_X$ is the unknown parallax error involved in the determination of the surface deviation $z$.

By introducing this result in Eq. \ref{eq:full system}, we get the final result
\begin{equation}
\begin{array}{c}
x=\frac{D_{xb}x_a-D_{xa}x_{b}}{D_{xb}-D_{xa}}\,+\,\frac{\mathrm{cos}(\phi)}{D_{xb}-D_{xa}}\epsilon_X+\,\frac{\mathrm{sin}(\phi)}{D_{xb}-D_{xa}}\epsilon_Y\\
y=\frac{D_{yb}y_a-D_{ya}y_{b}}{D_{yb}-D_{ya}}\,+\,\frac{\mathrm{sin}(\phi)}{D_{yb}-D_{ya}}\epsilon_X+\,\frac{\mathrm{cos}(\phi)}{D_{yb}-D_{ya}}\epsilon_Y\\\\
z=\frac{(D_{xb}-D_{xa})(x_{b}-x_{a})+(D_{yb}-D_{ya})(y_{b}-y_{a})}{(D_{xb}-D_{xa})^{2}+(D_{yb}-D_{ya})^{2}}\,+\,\frac{\epsilon_X}{[(D_{xb}-D_{xa})^2+(D_{yb}-D_{ya})^2]^{1/2}}
\end{array}
\label{eq:z_stereo}
\end{equation}
The errors are typically of the order of 1 pixel, corresponding to about 1 mm in our case. Since the wave slope is moderate, the error in $x$ and $y$ is of second order for the surface reconstruction. The expression of $z$ in \ref{eq:z_stereo} has been obtained by a linear combination of the two last equations of Eq. \ref{eq:full system} which eliminates $\epsilon_Y$. It gives therefore an optimum determination of $z$, and also  states how a probability distribution of $\epsilon_X$ translates into a probability distribution for the error in $z$.

The probability distribution of $\epsilon_X$ can be assumed identical with the distribution of the measured error $\epsilon_Y$, using an  hypothesis of isotropy. Then the error distribution of $z$ can be inferred from \ref{eq:z_stereo}. The error $\epsilon_Y$ can be also used  as a criterion to eliminate false data resulting form the image correlation procedure. This happens for instance with holes in the distribution of seeding particles. We know that the error in matching should not exceed a threshold of the order of one pixel, which can be translated into an error threshold for $\epsilon_Y$. This will be discussed further in section \ref{gravitywaves}.

\section{Stereoscopic PIV}
\label{PIVstereo}
%%%%%%%%%%%%%%

When particles are individually tracked, their velocity
can be obviously obtained by comparing their position $x,y,z$ at
successive times. However with the correlation method used in the stereo reconstruction, we only get
the position of the surface versus time, from which we can deduce
the normal velocity, not the full 3D velocity vectors. We thus perform image correlation from two successive images of each camera (PIV). Since the displacement between two successive images is rather small (unlike the displacement 
between the images of the two cameras considered previously), we avoid the step of image transform to the
apparent physical coordinates $(x_{a},y_{a})$, which is a source
of errors as it involves sub-pixel interpolation. We therefore obtain the displacements $dX_{a},dY_{a}$
in image coordinates on the sensor of camera \emph{a}. This is related
to the physical displacement by $dX_{a}=\frac{\partial F_{a}}{\partial x}dx+\frac{\partial F_{a}}{\partial y}dy+\frac{\partial F_{a}}{\partial z}dz$
and $dY_{a}=\frac{\partial G_{a}}{\partial x}dx+\frac{\partial G_{a}}{\partial y}dy+\frac{\partial G_{a}}{\partial z}dz$.
Comparing with the result of camera \emph{b} at the same physical point yields
in total 4 relations for the three unknown $dx,dy,dz$. This leads
to a solvability condition which is satisfied only in the absence
of experimental errors.
Therefore we replace the four equations by error estimates

\begin{equation}
\begin{array}{c}
\epsilon_{xa}=\frac{\partial F_{a}}{\partial x}dx+\frac{\partial F_{a}}{\partial y}dy+\frac{\partial F_{a}}{\partial z}dz-dX_{a}\\
\epsilon_{ya}=\frac{\partial G_{a}}{\partial x}dx+\frac{\partial G_{a}}{\partial y}dy+\frac{\partial G_{a}}{\partial z}dz-dY_{a}\\
\epsilon_{xb}=\frac{\partial F_{b}}{\partial x}dx+\frac{\partial F_{b}}{\partial y}dy+\frac{\partial F_{b}}{\partial z}dz-dX_{a}\\
\epsilon_{yb}=\frac{\partial G_{b}}{\partial x}dx+\frac{\partial G_{b}}{\partial y}dy+\frac{\partial G_{b}}{\partial z}dz-dY_{b}
\end{array}
\label{eq:epsPIV}
\end{equation}
and seek the physical displacement $(dx,dy,dz)$ which minimizes
the overall quadratic error $\epsilon_{xa}^{2}+\epsilon_{ya}^{2}+\epsilon_{xb}^{2}+\epsilon_{yb}^{2}$.
The corresponding optimum will give an estimate of the measurement
error. 

We have the partial derivatives
\begin{equation}
\begin{array}{c}
\frac{1}{2}\frac{\partial}{\partial(dx)}(\epsilon_{xa}^{2}+\epsilon_{ya}^{2}+\epsilon_{xb}^{2}+\epsilon_{yb}^{2})=\frac{\partial F_{a}}{\partial x}\epsilon_{xa}+\frac{\partial G_{a}}{\partial x}\epsilon_{ya}+\frac{\partial F_{b}}{\partial x}\epsilon_{xb}+\frac{\partial G_{b}}{\partial x}\epsilon_{yb}\\
\frac{1}{2}\frac{\partial}{\partial(dy)}(\epsilon_{xa}^{2}+\epsilon_{ya}^{2}+\epsilon_{xb}^{2}+\epsilon_{yb}^{2})=\frac{\partial F_{a}}{\partial y}\epsilon_{xa}+\frac{\partial G_{a}}{\partial y}\epsilon_{ya}+\frac{\partial F_{b}}{\partial y}\epsilon_{xb}+\frac{\partial G_{b}}{\partial y}\epsilon_{yb}\\
\frac{1}{2}\frac{\partial}{\partial(dz)}(\epsilon_{xa}^{2}+\epsilon_{ya}^{2}+\epsilon_{xb}^{2}+\epsilon_{yb}^{2})=\frac{\partial F_{a}}{\partial z}\epsilon_{xa}+\frac{\partial G_{a}}{\partial z}\epsilon_{ya}+\frac{\partial F_{b}}{\partial z}\epsilon_{xb}+\frac{\partial G_{b}}{\partial z}\epsilon_{yb}
\end{array}
\end{equation}
The condition of error minimization is obtained by setting to zero
each partial derivative, which yields a linear system of 3 equations
\begin{equation}
\begin{cases}
D_{11}dx+D_{12}dy+D_{13}dz=\frac{\partial F_{a}}{\partial x}dX_{a}+\frac{\partial G_{a}}{\partial x}dY_{a}+\frac{\partial F_{b}}{\partial x}dX_{b}+\frac{\partial G_{b}}{\partial x}dY_{b}\\
D_{21}dx+D_{22}dy+D_{23}dz=\frac{\partial F_{a}}{\partial y}dX_{a}+\frac{\partial G_{a}}{\partial y}dY_{a}+\frac{\partial F_{b}}{\partial y}dX_{b}+\frac{\partial G_{b}}{\partial y}dY_{b}\\
D_{31}dx+D_{32}dy+D_{33}dz=\frac{\partial F_{a}}{\partial z}dX_{a}+\frac{\partial G_{a}}{\partial z}dY_{a}+\frac{\partial F_{b}}{\partial z}dX_{b}+\frac{\partial G_{b}}{\partial z}dY_{b}
\end{cases}\label{eq:lin_system}
\end{equation}
with the symmetric matrix $(D_{ij}=D_{ji})$ defined by
\begin{equation}
\begin{array}{c}
D_{11}=(\frac{\partial F_{a}}{\partial x})^{2}+(\frac{\partial G_{a}}{\partial x})^{2}+(\frac{\partial F_{b}}{\partial x})^{2}+(\frac{\partial G_{b}}{\partial x})^{2}\\
D_{12}=\frac{\partial F_{a}}{\partial x}\frac{\partial F_{a}}{\partial y}+\frac{\partial G_{a}}{\partial x}\frac{\partial G_{a}}{\partial y}+\frac{\partial F_{b}}{\partial x}\frac{\partial F_{b}}{\partial y}+\frac{\partial G_{b}}{\partial x}\frac{\partial G_{b}}{\partial y}\\
D_{13}=\frac{\partial F_{a}}{\partial x}\frac{\partial F_{a}}{\partial z}+\frac{\partial G_{a}}{\partial x}\frac{\partial G_{a}}{\partial z}+\frac{\partial F_{b}}{\partial x}\frac{\partial F_{b}}{\partial z}+\frac{\partial G_{b}}{\partial x}\frac{\partial G_{b}}{\partial z}\\
D_{22}=(\frac{\partial F_{a}}{\partial y})^{2}+(\frac{\partial G_{a}}{\partial y})^{2}+(\frac{\partial F_{b}}{\partial y})^{2}+(\frac{\partial G_{b}}{\partial y})^{2}\\
D_{23}=\frac{\partial F_{a}}{\partial y}\frac{\partial F_{a}}{\partial z}+\frac{\partial G_{a}}{\partial y}\frac{\partial G_{a}}{\partial z}+\frac{\partial F_{b}}{\partial y}\frac{\partial F_{b}}{\partial z}+\frac{\partial G_{b}}{\partial y}\frac{\partial G_{b}}{\partial z}\\
D_{33}=(\frac{\partial F_{a}}{\partial z})^{2}+(\frac{\partial G_{a}}{\partial z})^{2}+(\frac{\partial F_{b}}{\partial z})^{2}+(\frac{\partial G_{b}}{\partial z})^{2}
\end{array}
\end{equation}

The displacements ($dx,dy,dz)$ are then obtained as solution of the
linear system Eq. \ref{eq:lin_system}, from which the corresponding velocity
components are deduced by division by the time interval $dt$. The coefficients of the equations are known from the calibration functions (as specified in the appendix) and the position $x,y,z$ obtained by the stereoscopic surface mapping described in the previous section. 

The method gives also an error estimate
\begin{equation}
\epsilon'=\frac{1}{2}(\epsilon_{xa}^{2}+\epsilon_{ya}^{2}+\epsilon_{xb}^{2}+\epsilon_{yb}^{2})^{1/2}\label{eq:error_PIV}
\end{equation}
which is expressed in pixel displacement.

\section{Application to surface gravity waves}
\label{gravitywaves}

The PIV-Stereo method has been developed as part of the ERC-funded WATU project for experimental investigations of non-linear wave interaction in the framework of the weak wave turbulence theory. It has been successfully deployed on the Coriolis facility (LEGI, Grenoble) for a fully-resolved measurement of surface gravity waves as well as for internal gravity waves. The Coriolis facility is used as a large cylindrical wave tank 13~m in diameter filled with fresh water at a rest height of 70 cm. Waves are produced by two triangular wedges undergoing vertical oscillation at frequency randomly fluctuating around a reference value.

The surface is seeded with polystyrene beads $700$ $\mu m$ in diameter used by factories to produce expanded polystyrene. Those contain pentane gas which expands by heating like pop corn. We use a moderate heating which provides a density a little lower than 1 (about 0.95). Then the particles are floating but they are well anchored in water. They are not swept by air flow perturbations like observed by \cite{weitbrecht2002} for fully expanded polystyrene particles. 

\begin{figure}
\includegraphics[width=0.7\textwidth]{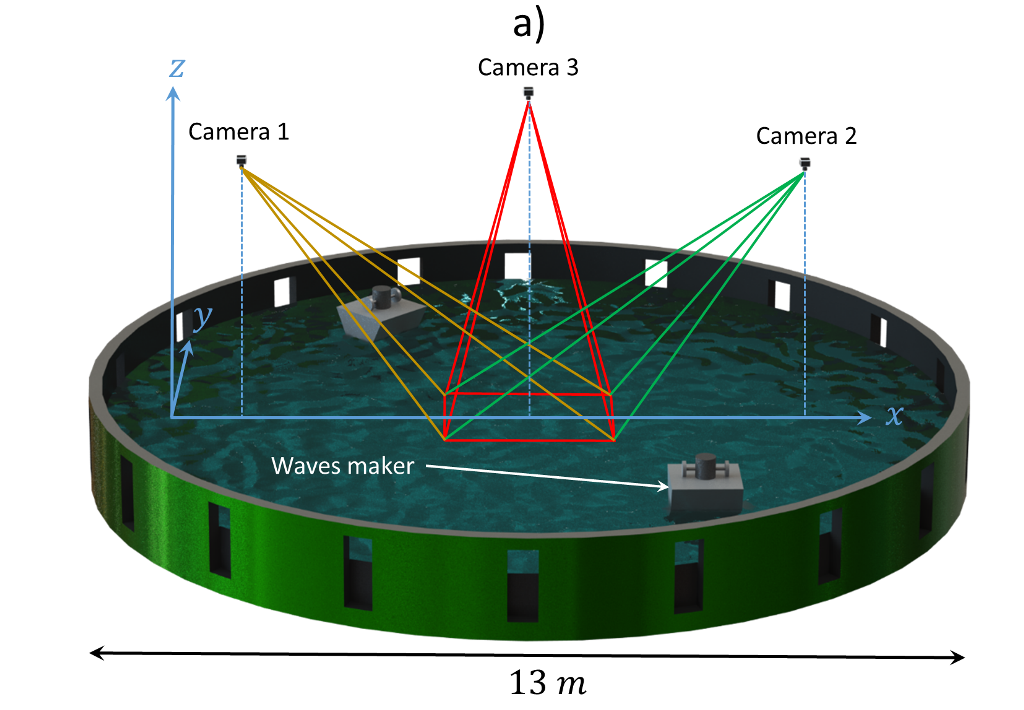}
\includegraphics[width=0.29\textwidth]{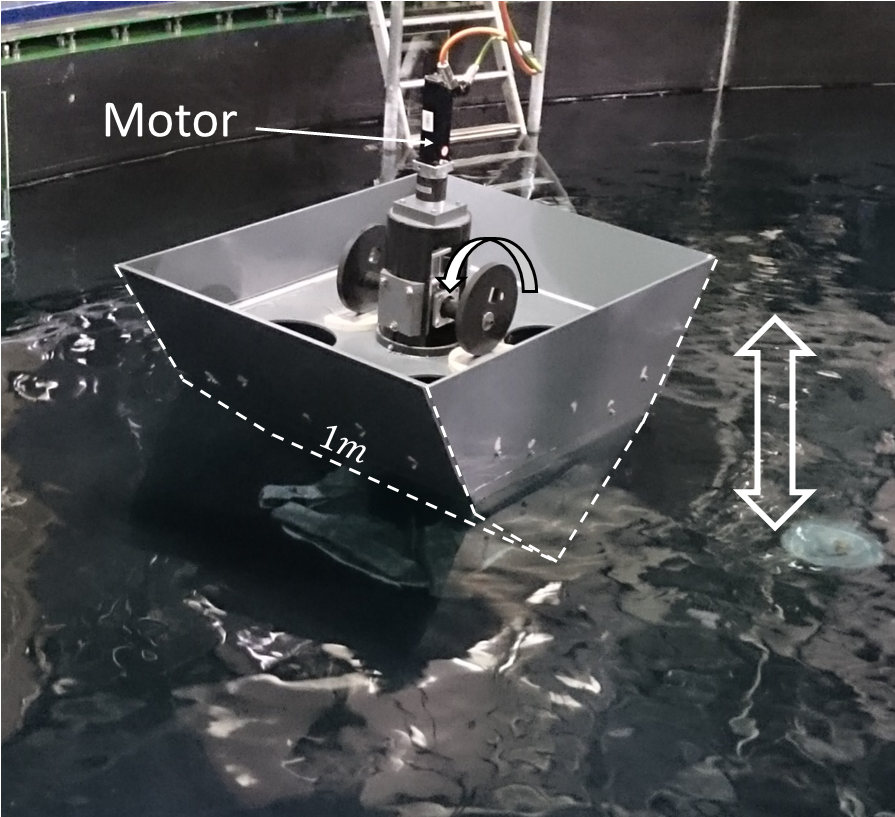}
\includegraphics[height=6.5cm]{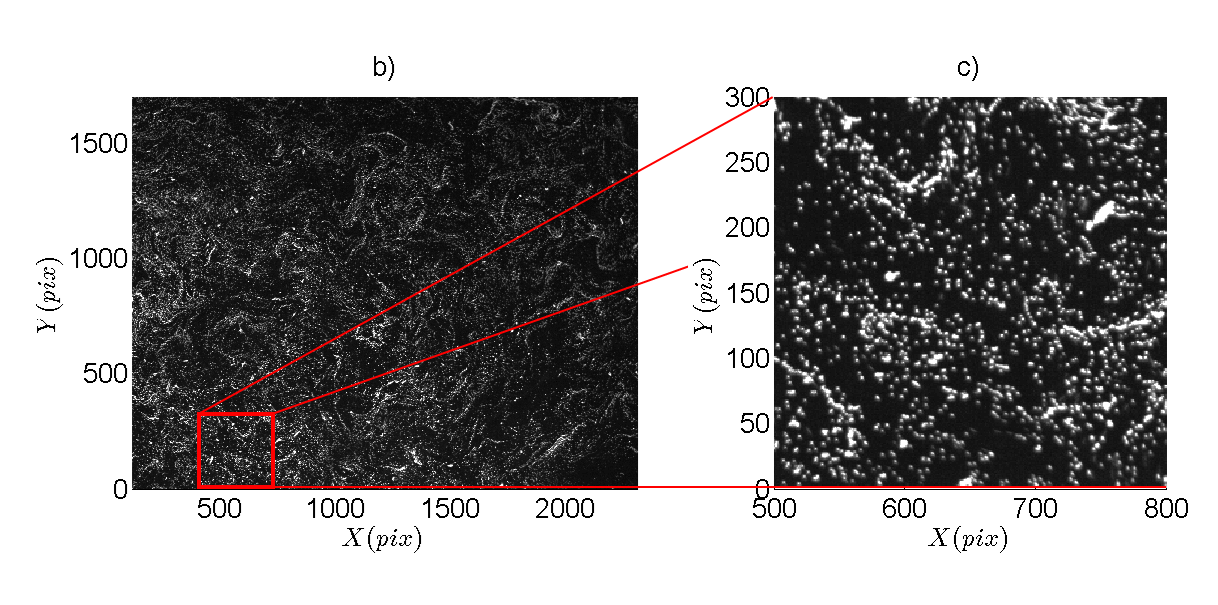}
\caption{a) Diagram of the experiment performed on the Coriolis facility (LEGI,
Grenoble France). The tank is filled with water at a rest height
of $70$~cm. Two vertically oscillating wave-makers, as shown on the right side,  are used to generate a homogeneous gravity
wave field. Three cameras are placed on the top to perform a fully-resolved
measurement using the PIV-Stereo method over a surface of $1.5\times2$~m$^{2}$.
b) Snapshot of the top camera 3 from a running experiment. Particles at the
surface are illuminated through six halogen projectors placed at the
edge of the tank. Waves of a few centimeters in amplitude are
present and recorded both by cameras and local capacitive probes.c) zoom on the particles showing the quality of the seeding.
\label{fig:schema_manip}}
\end{figure}

 An ensemble of three cameras is fixed at roughly 4 m on top of the free surface as visualized in Fig. \ref{fig:schema_manip}(a). The three cameras are aligned along the $x$ direction oriented $45{^\circ}$ apart, such that the camera 3 at the middle is normal to the surface. The view field common to the three cameras is about $1.5\times2$~m$^{2}$. The three cameras are mounted with a Nikon lens with a $35$~mm focal length. The lens of cameras 1 and 2 are hold by two homemade Scheimpflugs mounts to improve the depth of field by compensating the cameras tilt with respect to the surface.
 
The camera 1 and 2 have a resolution $1024^{2}$ pixels (trade mark Dalsa Panthera 1M60), while the central camera 3 has resolution $2432\times1728$ pixels (trade mark Falcon 4M). The corresponding spatial resolution is thus imposed by the lowest resolution of cameras 1 and 2, 1.5 mm in $y$ and 2 mm in $x$  due to the projection effect, so the mean spatial resolution can be estimated as 1.8 mm. The three cameras are synchronized by TTL electrical pulses provided by a commercial system (RG from RD Vision), and the images are recorded at a frame rate of 20 Fps ($ dt=0.05$~s). Typically a series of 12~000 frames (per camera) is performed, providing  wave statistics from a continuous record of 10 minutes.

Particles are illuminated with six 1 kW halogen projectors through windows located in the vertical outer rim of the tank. The large incidence angle of the light on the free surface leads to total reflection, which avoids the direct illumination of
the camera sensor. Fig. \ref{fig:schema_manip}(b)
shows a snapshot of the top camera 3 during an acquisition. Particles are well dispersed although they cluster in patterns caused by the chaotic waves field. Waves are generated with two wedge wave-makers vertically oscillating at 1 Hz with a typical amplitude of about $6$~cm. As visible in the sketch of Fig. \ref{fig:schema_manip}(a), they are placed near the wall and generate a well-mixed and homogeneous wave field. Four capacitive probes are also placed at the edge of the view field to get local high resolution measurements of the interface as a reference for the stereo imaging measurement. 

Our geometric calibration relies on a pin-hole camera model whose parameters are obtained by the classical procedure initiated by Tsai \cite{Tsai1987}. The small geometric aberrations induced by the lens are corrected with a quadratic deformation. For a practical implementation, we use a `camera calibration toolbox' provided online by Jean-Yves Bouguet (Caltech), which relies on  improvements to the Tsai's method brought by Heikkila et al. \cite{Heikkila1997}  and  Zhang \cite{Zhang1999}. The details of the implementation for the PIV-Stereo method are given in appendix \ref{Appendix}.

The stereoscopic cross-correlation is performed using pyramidal refinement (see for instance \cite{Scarano1999}).
The first step resolves the large scales by using a correlation box of about $25\times 25$~cm$^2$ ($125 \times 155$  pix$^2$) within a search windows of $35 \times 35$ cm$^{2}$($175 \times 220$ pix$^2$). The last step is limited by the seeding of particles which gives a spatial resolution of $4 \times 4$ cm$^2$($20 \times 25$ pix$^2$).
Using the cameras 1 and 2 with viewing  angle  nearly $90{^\circ}$  apart, the apparent horizontal displacement is roughly twice the vertical displacement. Thus for strong wave amplitudes, the cross-correlation may become difficult to obtain due to large displacement and some geometric distortions that appear during the interpolation in the reference plane.
This issue is fixed by using the camera 3 as intermediate. The cross-correlation for the stereo is then computed with the two pairs of camera $(1-3)$ and $(2-3$, with viewing angles separated by an angle close to $45{^\circ}$ . The two displacements are then added to get the apparent displacement $(x_a-x_b, y_a -y_b)$  with the pair $(1-2)$. Stereoscopic PIV is  performed with the camera pair $[1,2]$ which are 90$°$ apart to keep the best vertical sensitivity (see next section \ref{Tests_precision}).
Both PIV and stereo are finally merged as explained previously to obtain a full measurement of the interface: $z(x,y,t)$, $U(x,y,t)$,
$V(x,y,t)$ and $W(x,y,t)$. These fields are finally interpolated on a regular grid in $(x,y)$ (with mesh 1 cm) by a thin plate spline method (see appendix). Fig. \ref{fig:3D_Snapshot} (a) shows a 3D snapshot of the reconstructed free surface and b) the corresponding velocity field. 

\begin{figure}
\includegraphics[height=6.5cm]{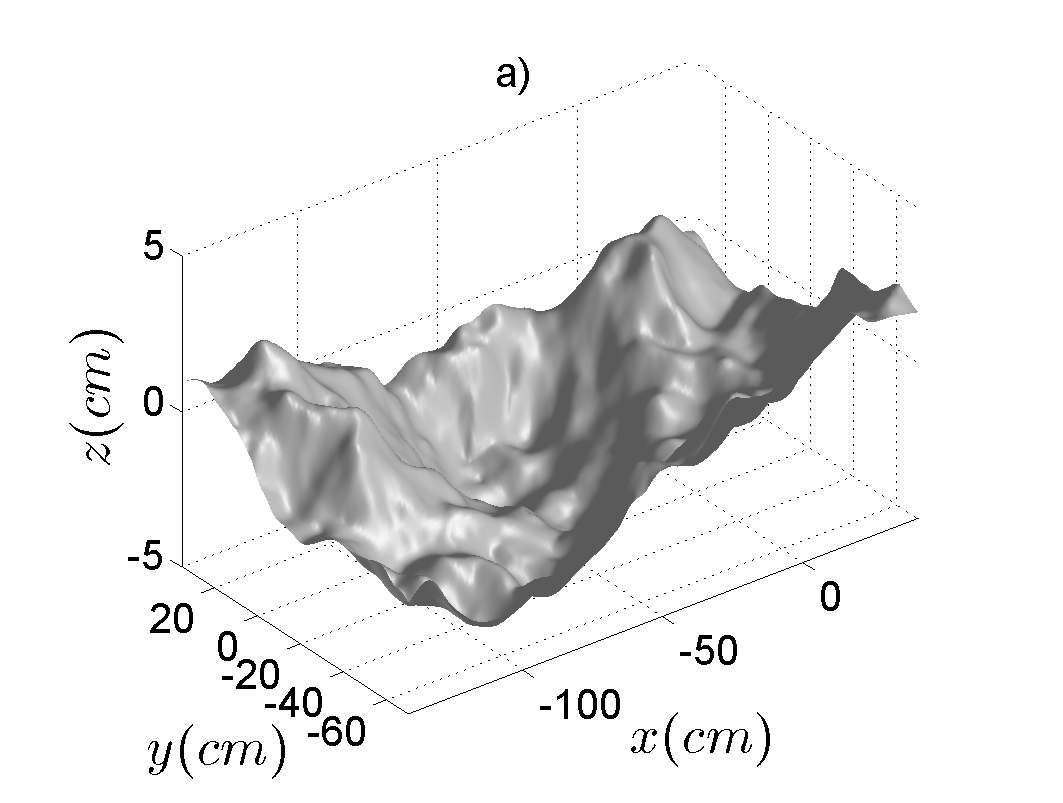}
\includegraphics[height=6.5cm]{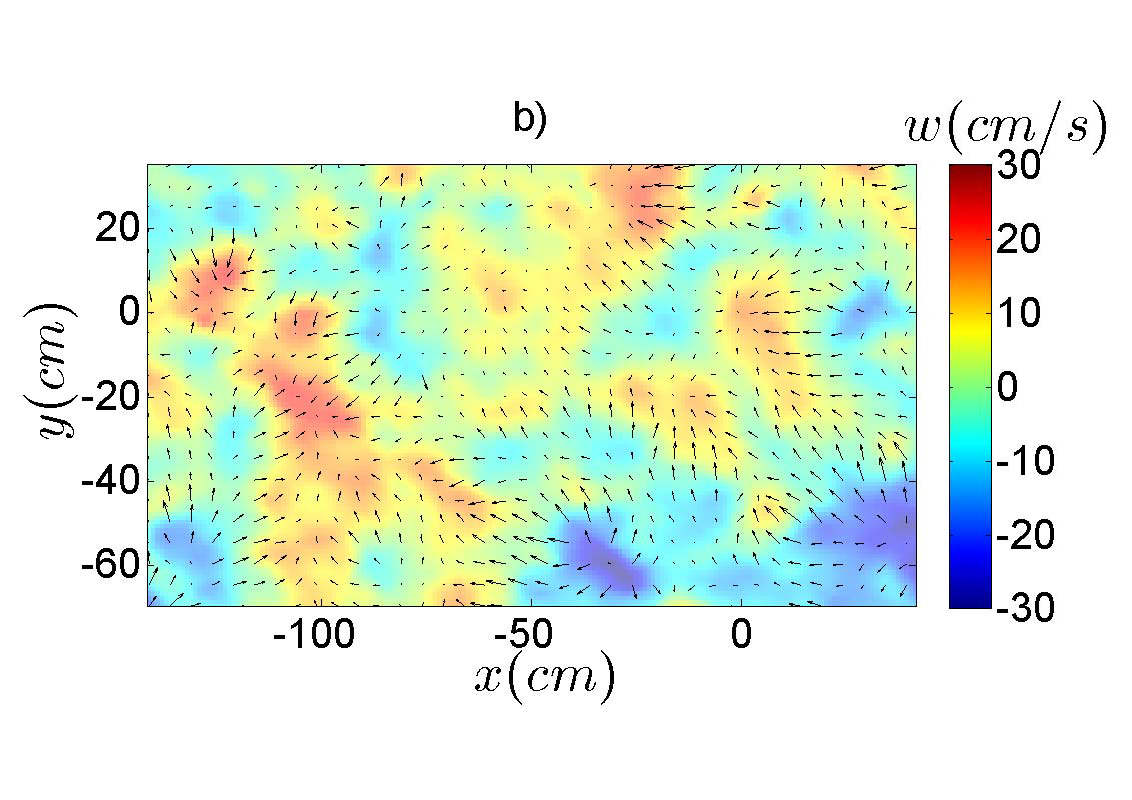}

\caption{a) 3D Snapshot of the free surface $z(x,y)$. A mix of waves of about
2 cm in maximum amplitude is visible. b) Corresponding velocity
field . Vertical velocity $w(x,y)$ is color coded. The
horizontal velocity $\left[u(x,y),v(x,y)\right]$
is plotted as vectors. \label{fig:3D_Snapshot}}
\end{figure}

\begin{figure}
\includegraphics[height=6cm]{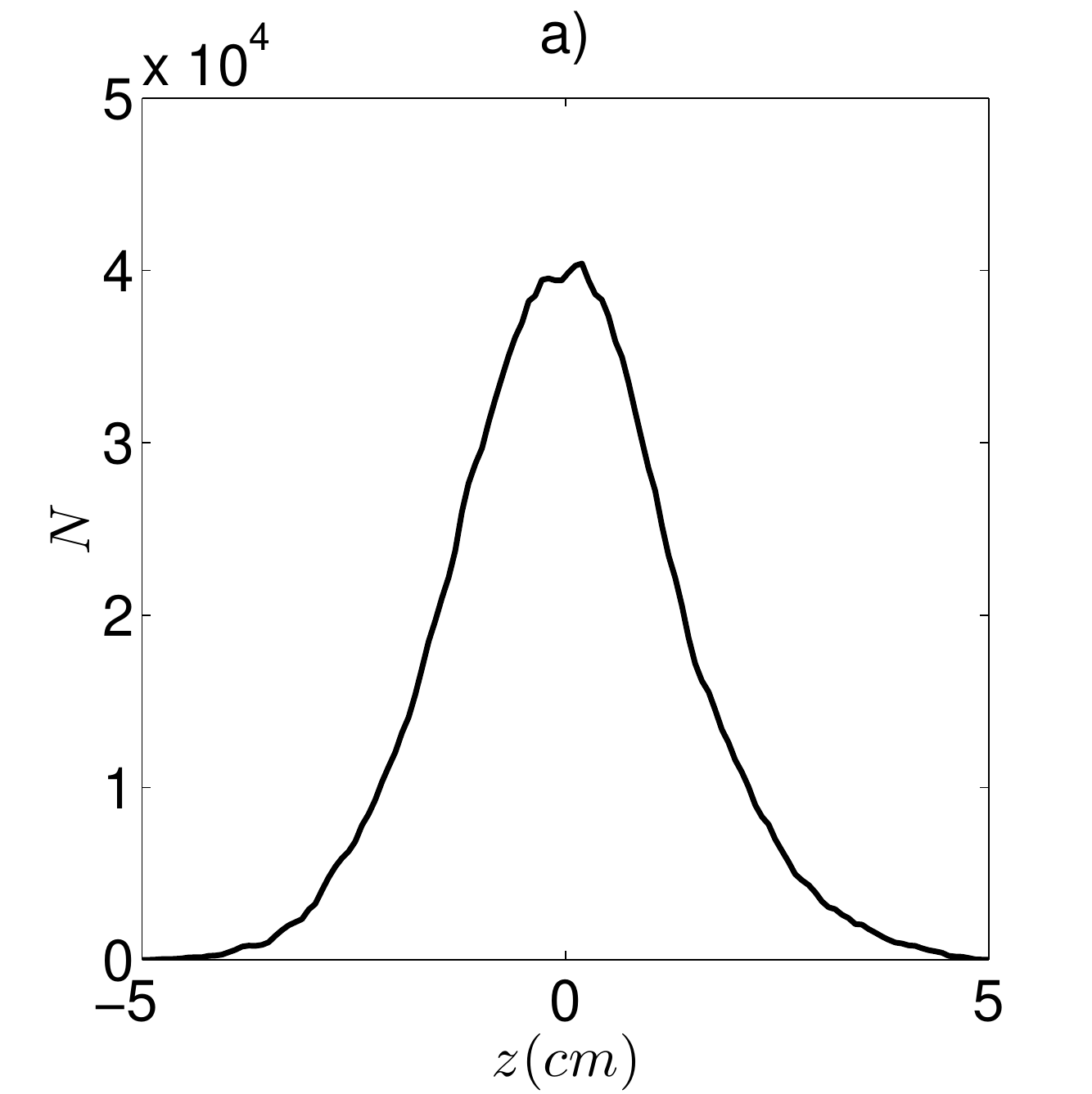}
\includegraphics[height=6cm]{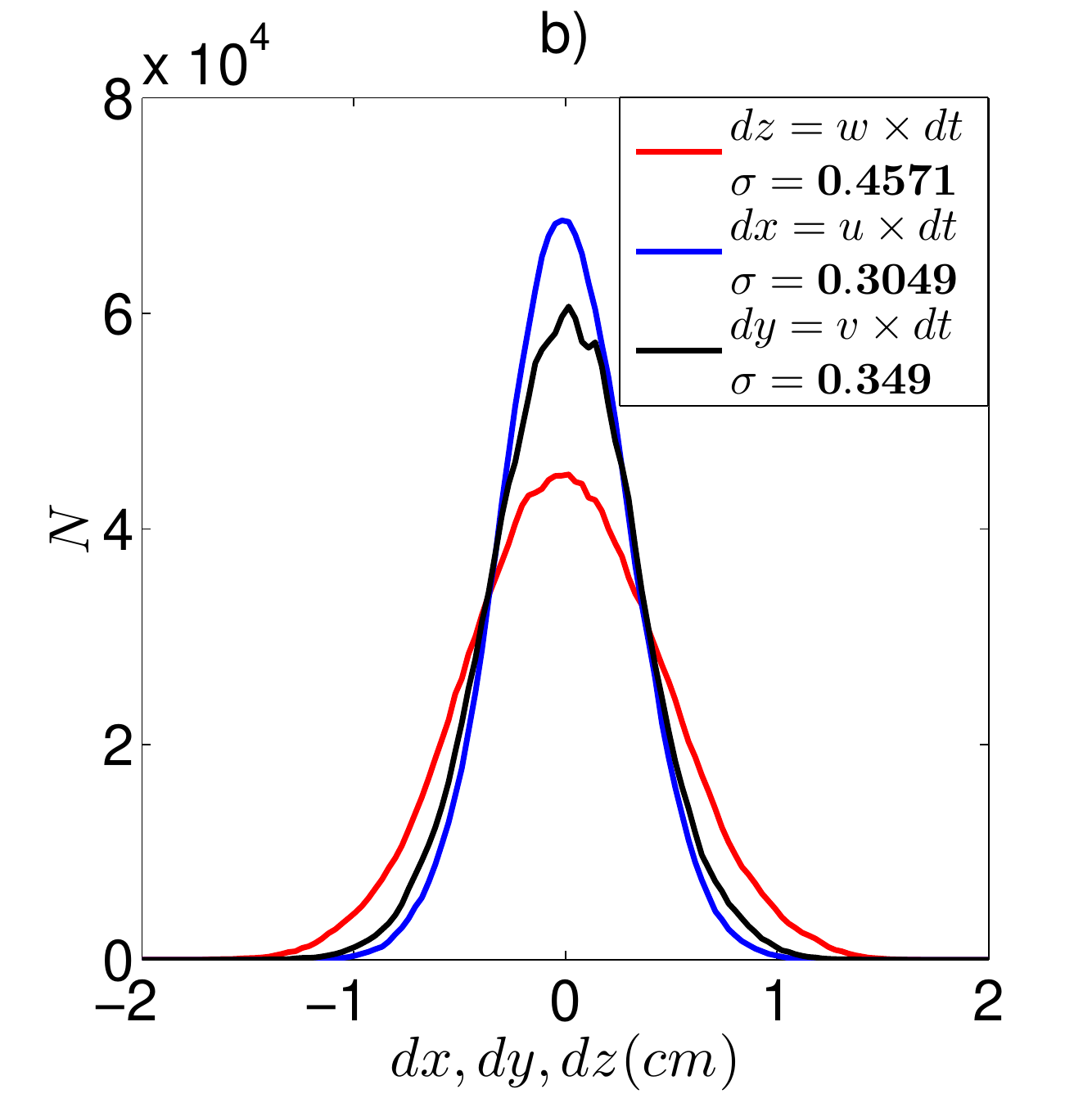}
\caption{a) Distribution of $z$ for one typical experiment. The maximum amplitude
of waves is about $5$~cm. b) Distribution of $dx=u\, dt$,$dy=v\, dt$ and $dz=w\, dt$ corresponding to a) ($dt=0.05$~s). $\sigma$ is the variance of the corresponding displacement. The vertical displacements $dz$ are slightly
larger than horizontal components. \label{fig: PDFHUVW} }
\end{figure}

We observe a superposition of different wave lengths in our field. Fig. \ref{fig: PDFHUVW} shows the probability distribution functions (pdf) of $z$ and the material displacements $dx=u\times dt$, $dy=v\times dt$ and $dz=w\times dt$, obtained for a time series of 10 minutes. 

\begin{figure}
\includegraphics[width=7cm]{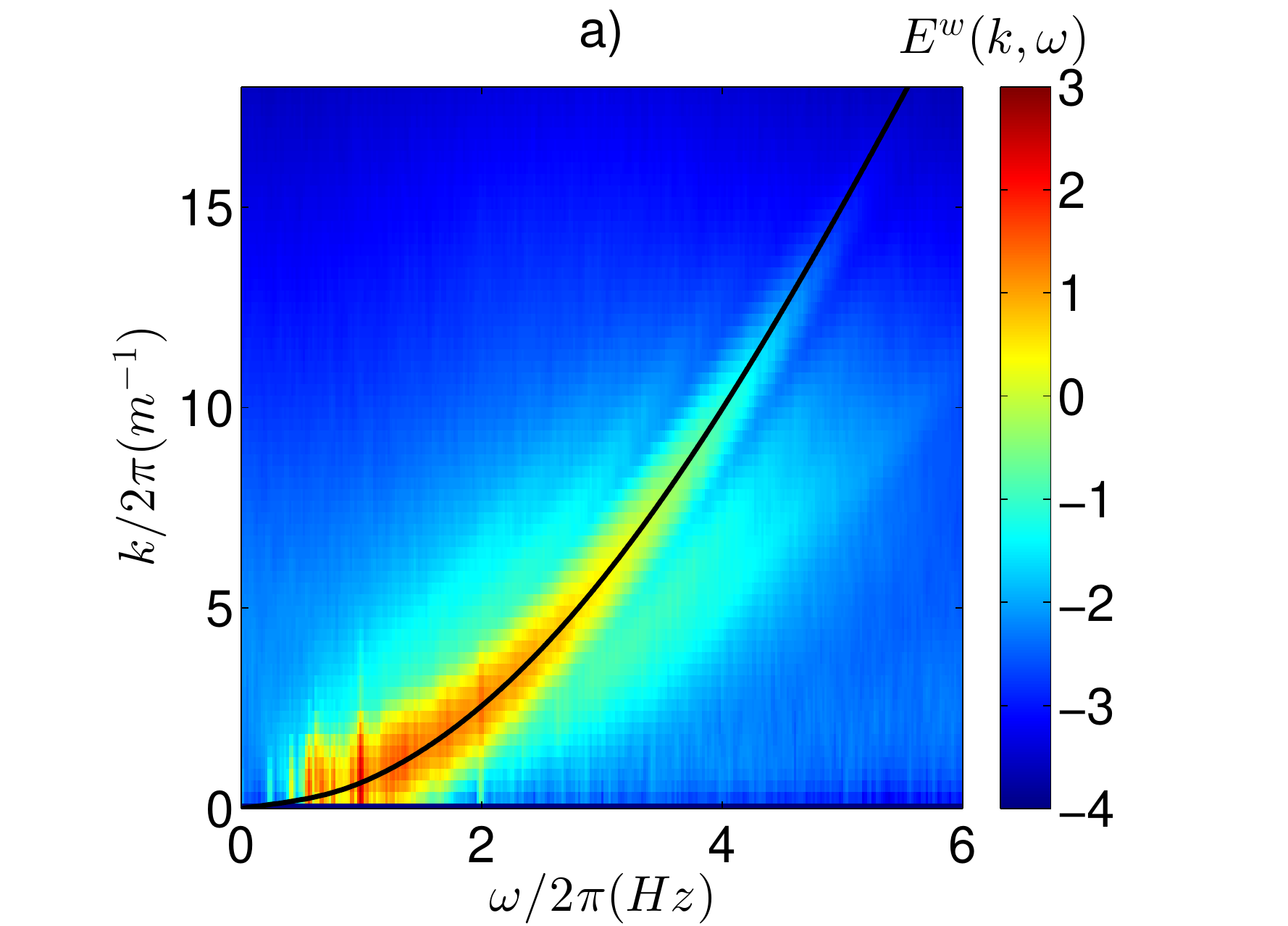}
\includegraphics[width=7cm]{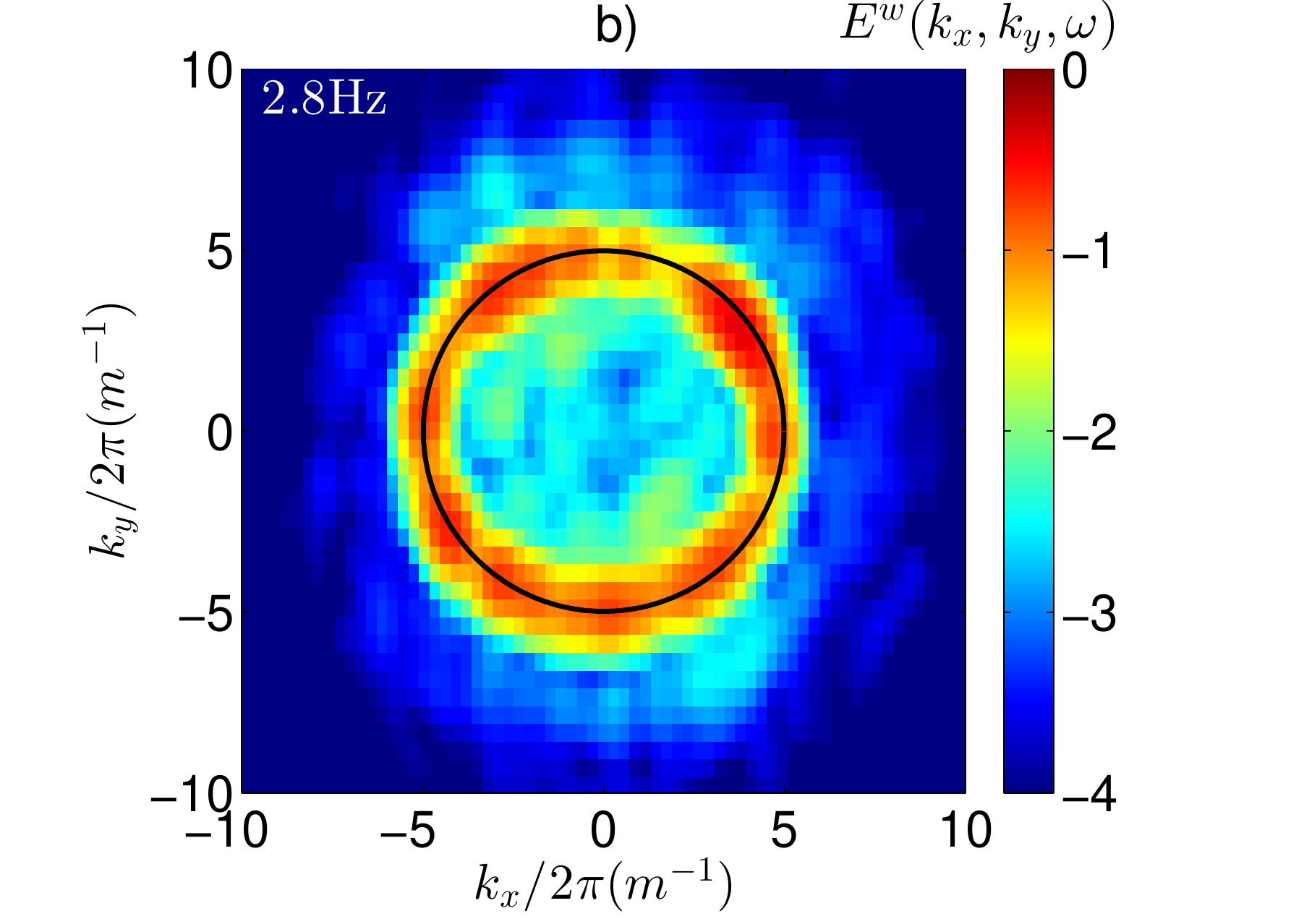}
\caption{a)Spatio-temporal power spectrum $E^{w}(k,\omega)$ of the surface
gravity waves (color coded  log$_{10}$ scale). The black line represents the
linear dispersion of the waves: $\omega=\sqrt{gk}$. The concentration
of the energy along this line confirms the measurement of waves at frequencies up
to $5$~Hz. The dynamic range of the measure in energy is near 5 orders
of magnitude, which corresponds to 2.5 orders for the velocity scale. b)
$E^{w}(\mathbf{k},\omega)$ at $\omega\text{/2\ensuremath{\pi}=2.8 Hz}.$ We
observe an isotropic  distribution of the waves in the horizontal plane. \label{fig:Spatio-temporal-energy}}
\end{figure}

We observe a quasi-Gaussian distribution of vertical displacements with maximum amplitude about $5$~cm. The vertical speed displacements $dz=w\, dt$ are slightly larger than horizontal components with displacements up to 1.5~cm. 
For waves, a proper way to estimate the noise and the dynamic range of the measure is to compute the spatio-temporal power spectrum $E^{w}(\mathbf{k},\omega)$. This is obtained from the 2D space and time Fourier transform $\hat{w}(\mathbf{k},\omega)$ of  the vertical velocity field $w(x,y,t)$, where $\mathbf{k}$ is the wave-vector and $\omega$ the frequency. We take an average of the modulus squared $\left\langle\left|\hat{w}(\mathbf{k},\omega)\right|^2\right\rangle$, from which the spectrum  $E^{w}(k,\omega)$  is obtained by an angular integration for each wavenumber $k=\left|\mathbf{k}\right|$. 
\begin{equation}
E^{w}(k,\omega)=\left\langle \left|\hat{w}(\mathbf{k},\omega)\right|^{2}\right\rangle 
\end{equation}
This spectrum is mapped in Fig. \ref{fig:Spatio-temporal-energy}. 

The spatio-temporal spectrum shows an energy concentration along the linear dispersion relation of surface gravity waves $\omega=\sqrt{gk}$ (black line).
This confirms the measurements of weakly non-linear waves up to roughly $5$~Hz in frequency (corresponding wavelength $6$~cm). 
Fig. \ref{fig:Spatio-temporal-energy} b) shows the spatial distribution of the wave energy at a given frequency $\omega/2\pi=2.8$~Hz.
The system appears to be fairly isotropic. The range of energy observable reaches 5 orders of magnitude, which corresponds approximately to 2.5 orders or magnitude for
the velocity amplitude.

The knowledge of the three velocity components allows us to directly measure the non-linearities in this system. The vertical velocity of the free surface deformation $\partial \eta/\partial t$ is indeed related to $w$ through the equation 
\begin{equation}
\frac{\partial \eta}{\partial t}=w-\mathbf{u}\cdot\nabla \eta 
\label{equ:vitesse}
\end{equation}
with $\mathbf{u}=(u,v)$ the horizontal velocity vector and $\nabla \eta$ the horizontal gradient of the surface deviation $z=\eta(x,y,t)$. The advective term $-\mathbf{u}\cdot\nabla \eta$ thus represents the non-linearities.

\begin{figure}
\includegraphics[width=7cm]{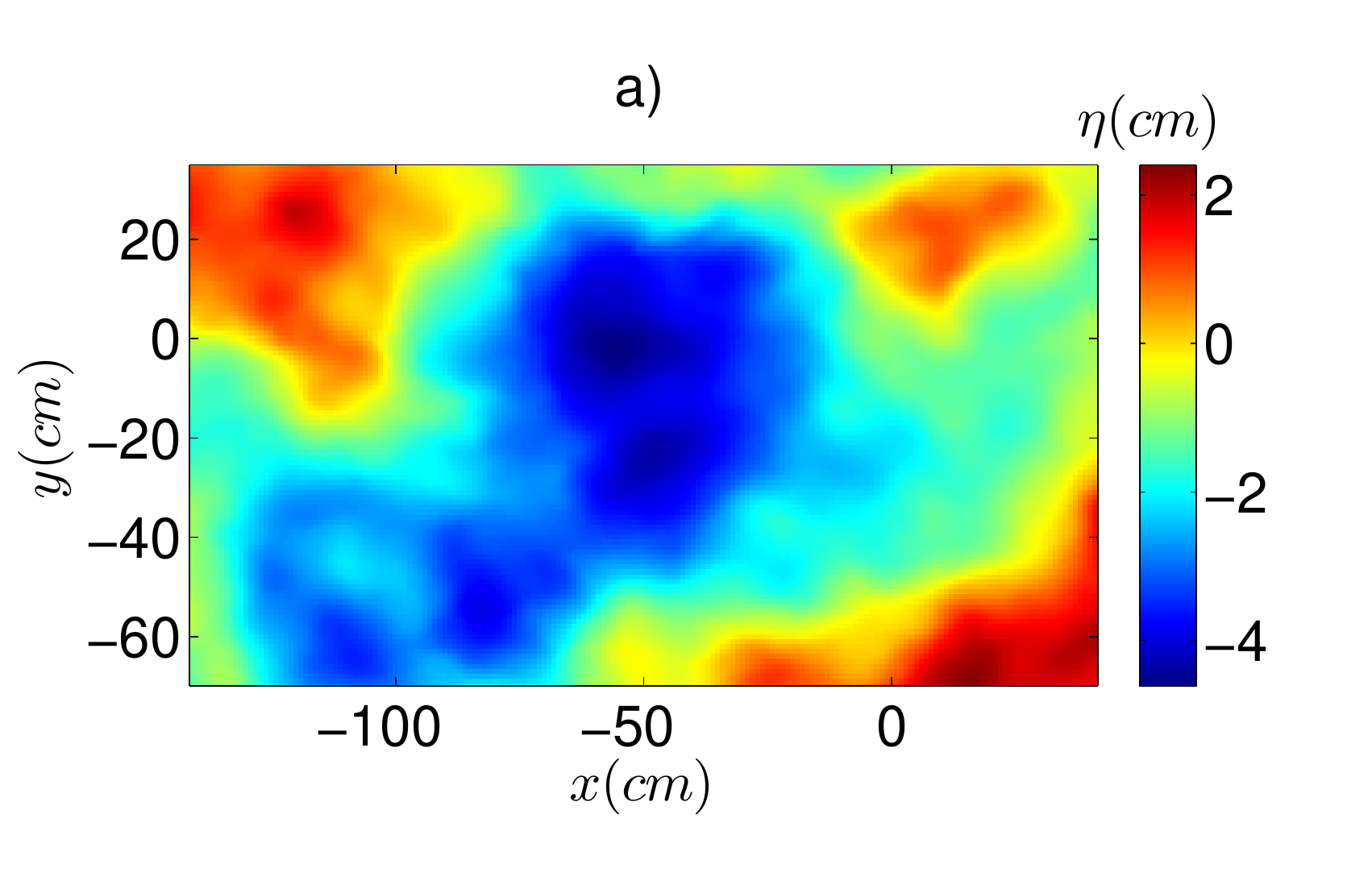}
\includegraphics[width=7cm]{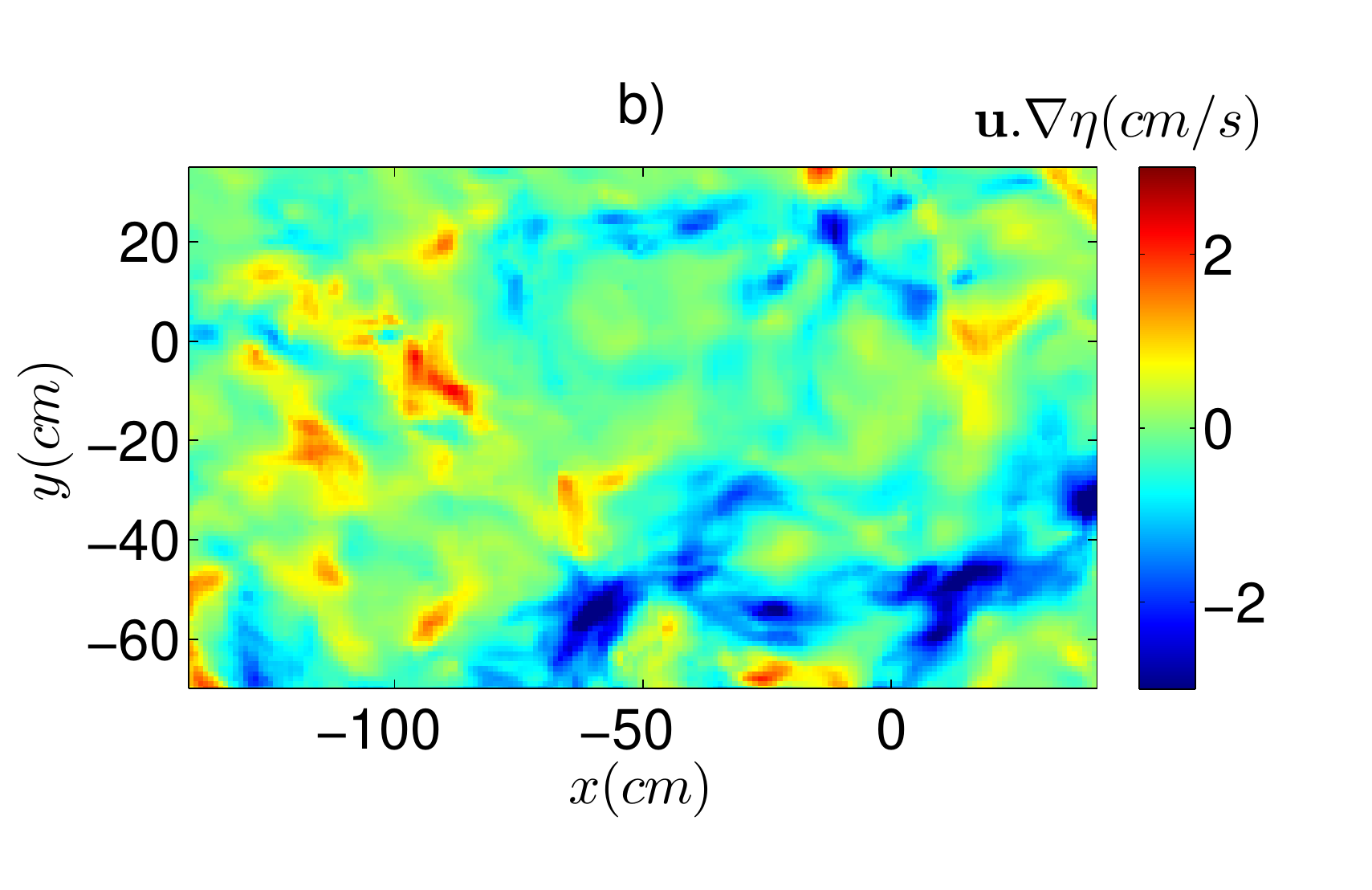}
\includegraphics[width=7cm]{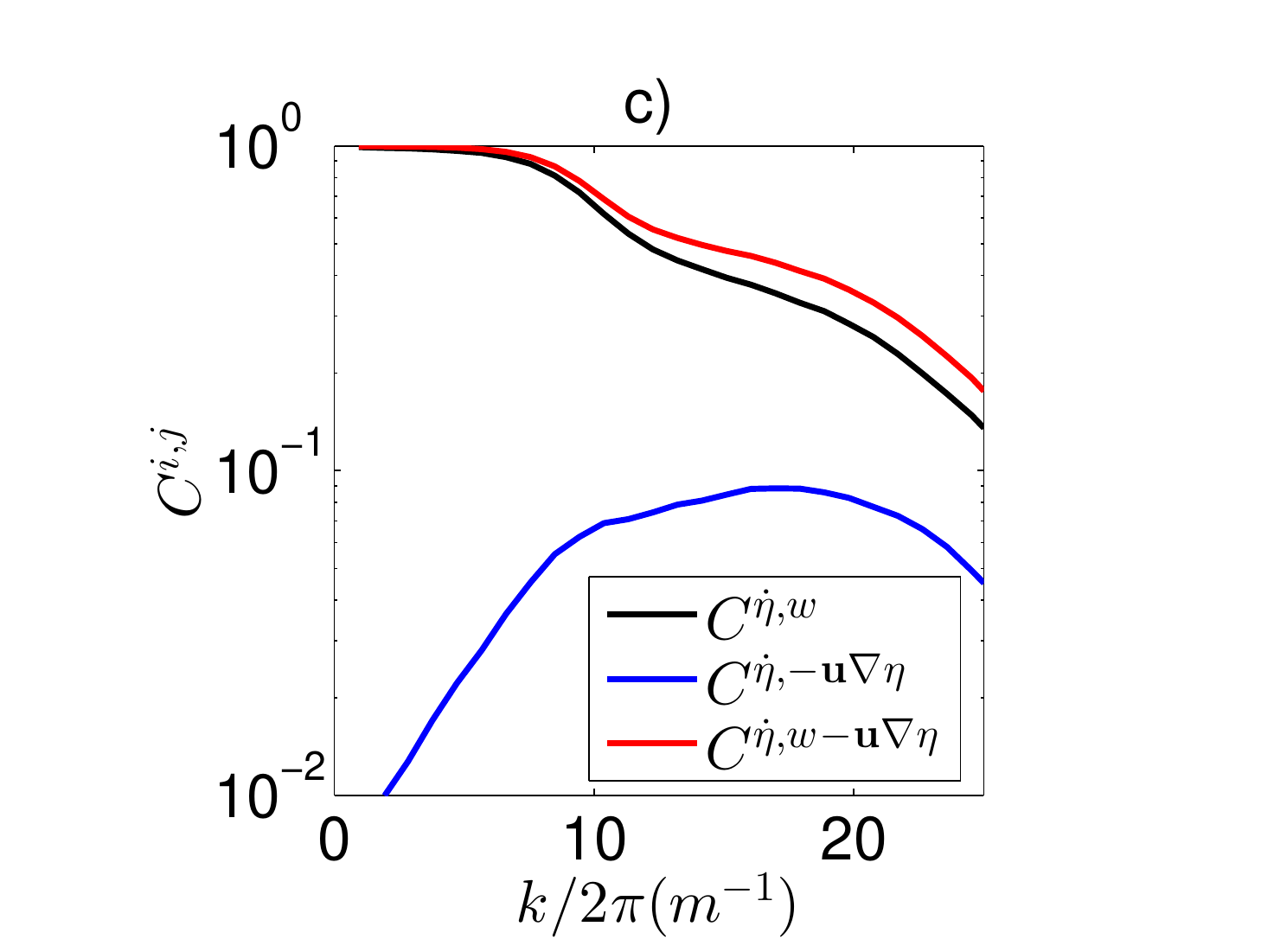}
\includegraphics[width=7cm]{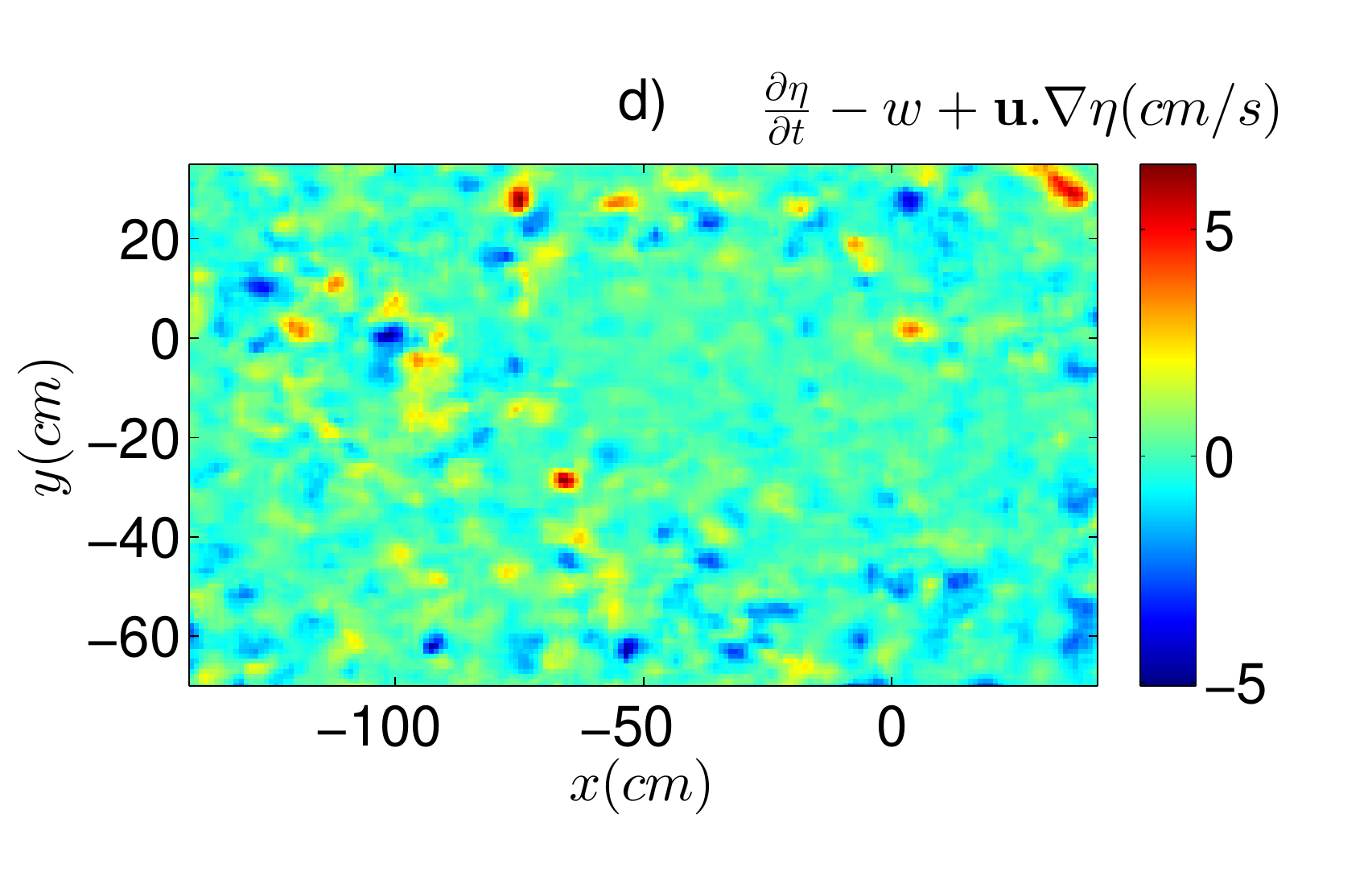}
\caption{a) Color map of the vertical surface displacement $z=\eta(x,y)$ ( also represented  in Fig. \ref{fig:3D_Snapshot} a). b)Value of the non linear term $\mathbf{u}.\nabla \eta$ at the same time . b) Coherence level along the $k$ dimension over a time series of 10 minutes. The black curve is the coherence between the vertical velocity of surface deformation $\dot{\eta}\equiv\partial\eta/\partial t$ and the vertical velocity of the fluid $w$ at the same position. The blue curve is the coherence between $\dot{\eta}$ and $-\mathbf{u}\nabla \eta$. The red curve is the coherence between $\partial{\eta}/partial{t}$ and $w-\mathbf{u}.\nabla \eta$.}. d) Map of the error field $\partial{\eta}/partial{t}-w+\mathbf{u}.\nabla \eta$ at the same time as a) and b).
\label{fig:nonlin}
\end{figure}

Fig. \ref{fig:nonlin}b displays a snapshot of this non-linear term at the same time as the deformation $\eta(x,y)$ shown in Fig. \ref{fig:nonlin}a (also represented in Fig. \ref{fig:3D_Snapshot}a. This is an order of magnitude smaller than the vertical velocity shown in Fig. \ref{fig:3D_Snapshot}b. The error obtained by subtracting the two members of Eq. \ref{equ:vitesse} is of the same order as the nonlinear term, see Fig. \ref{fig:nonlin}d,  but it is limited to small scale noise. A statistical description of these quantities is obtained by the computation of the spectral coherence $C$ between the different terms of Eq. \ref{equ:vitesse} defined as:
\begin{equation}
\begin{array}{c}
C^{\dot{\eta},w}(\mathbf k)=\frac{\left|\left\langle  \dot{\eta}^*(\mathbf k,t)w(\mathbf k,t)\right\rangle\right|}{\sqrt{\left\langle \left| \dot{\eta}\right|^2\right\rangle \left\langle \left| (w-\mathbf{u}.\nabla \eta) \right|^2\right\rangle}}\\
C^{\dot{\eta},-\mathbf{u}.\nabla \eta}(\mathbf k)=\frac{\left|\left\langle  \dot{\eta}^*(\mathbf k,t)(-\mathbf{u}.\nabla \eta)(\mathbf k,t)\right\rangle\right|}{\sqrt{\left\langle \left| \dot{z}\right|^2\right\rangle \left\langle \left| (w-\mathbf{u}.\nabla \eta) \right|^2\right\rangle}}\\
C^{\dot{\eta},w-\mathbf{u}.\nabla \eta}(\mathbf k)=\frac{\left|\left\langle  \dot{\eta}^*(\mathbf k,t)(w-\mathbf{u}.\nabla \eta)(\mathbf k,t)\right\rangle\right|}{\sqrt{\left\langle \left| \dot{\eta}\right|^2\right\rangle \left\langle \left| (w-\mathbf{u}.\nabla \eta) \right|^2\right\rangle}}
\end{array}
\label{eq:coh}
\end{equation}
The average $\langle \cdot \rangle$ is an average over time. Using our choice of a common normalization of the coherence allows us to compare directly the three coherence estimators. The coherence $C^{\dot{\eta},w-\mathbf{u}.\nabla \eta}$ should be equal to 1 in principle. It is the case at low $k$ value but it decays at large $k$ due to the presence of measurement noise as shown in Fig. \ref{fig:nonlin}d (noise becomes dominant beyond $k/(2\pi)$=20~m$^{-1}$ corresponding to $\omega/(2\pi)=6$~Hz, which fits with time spectra of Fig. \ref{fig:Temporal-spectrum-}. We observe that the coherence fraction of the non-linear term increases with $k$ up to a maximum around $10$\%. The proximity of the black and red curves confirm the weakly-nonlinear character of our system which is thus expected to be in the range of validity of the Weak Turbulence Theory.

\section{Accuracy of the method}
\label{Tests_precision}

\begin{figure}
\includegraphics[width=0.49\textwidth]{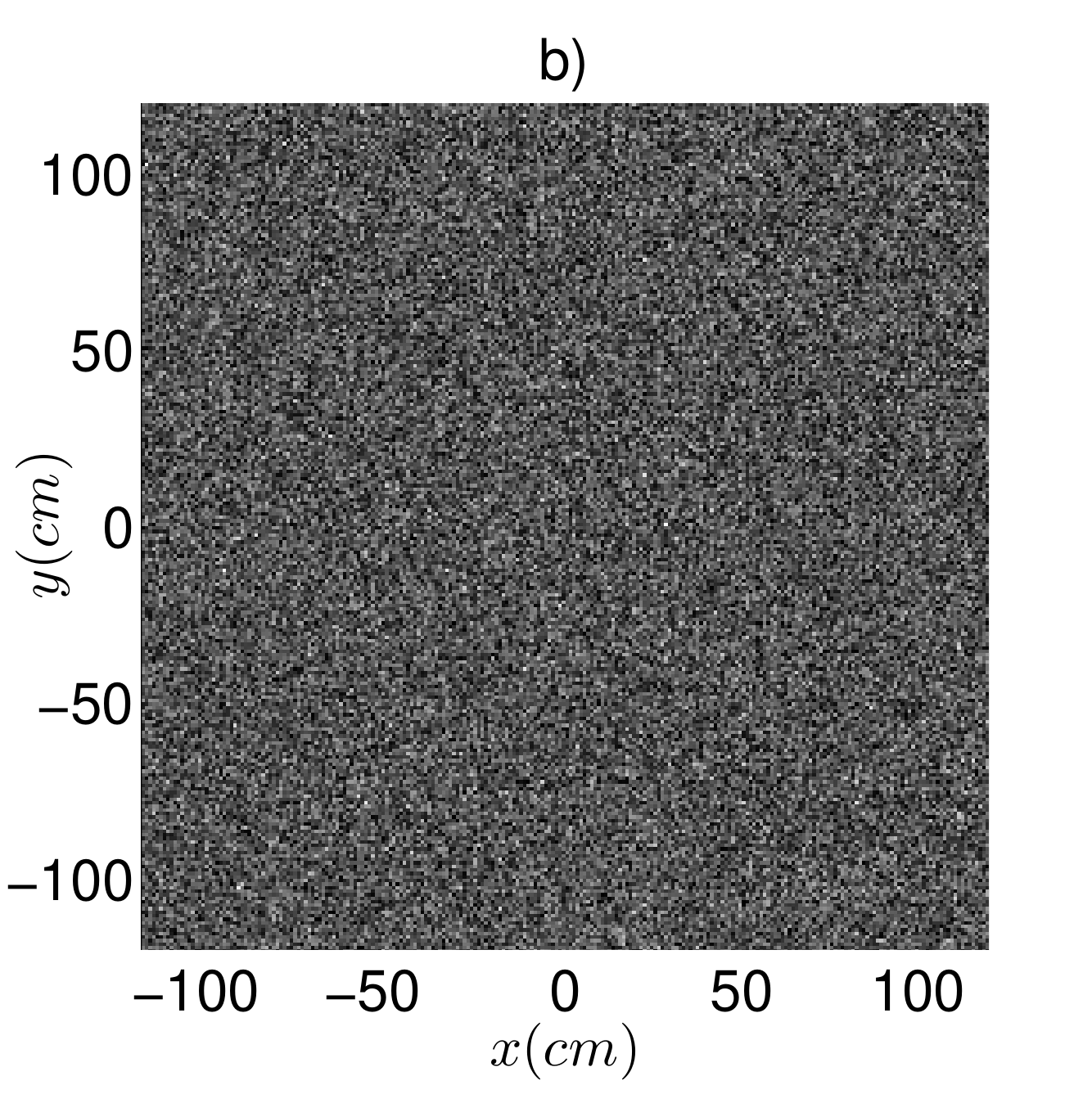}
\includegraphics[width=0.49\textwidth]{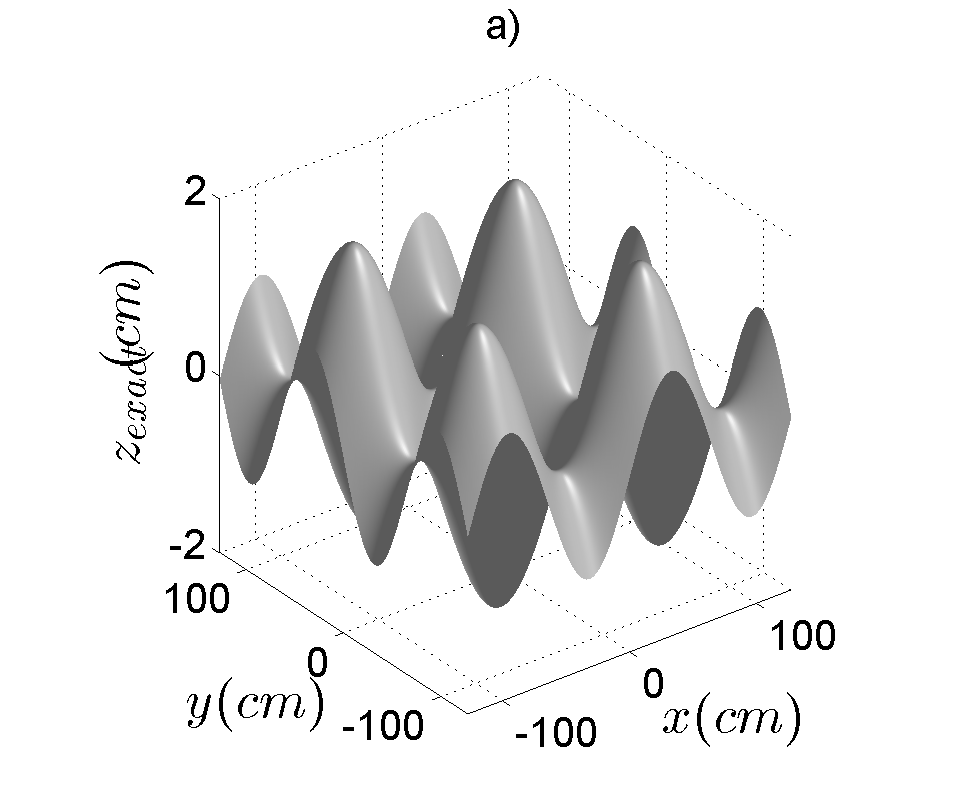}
\includegraphics[width=0.49\textwidth]{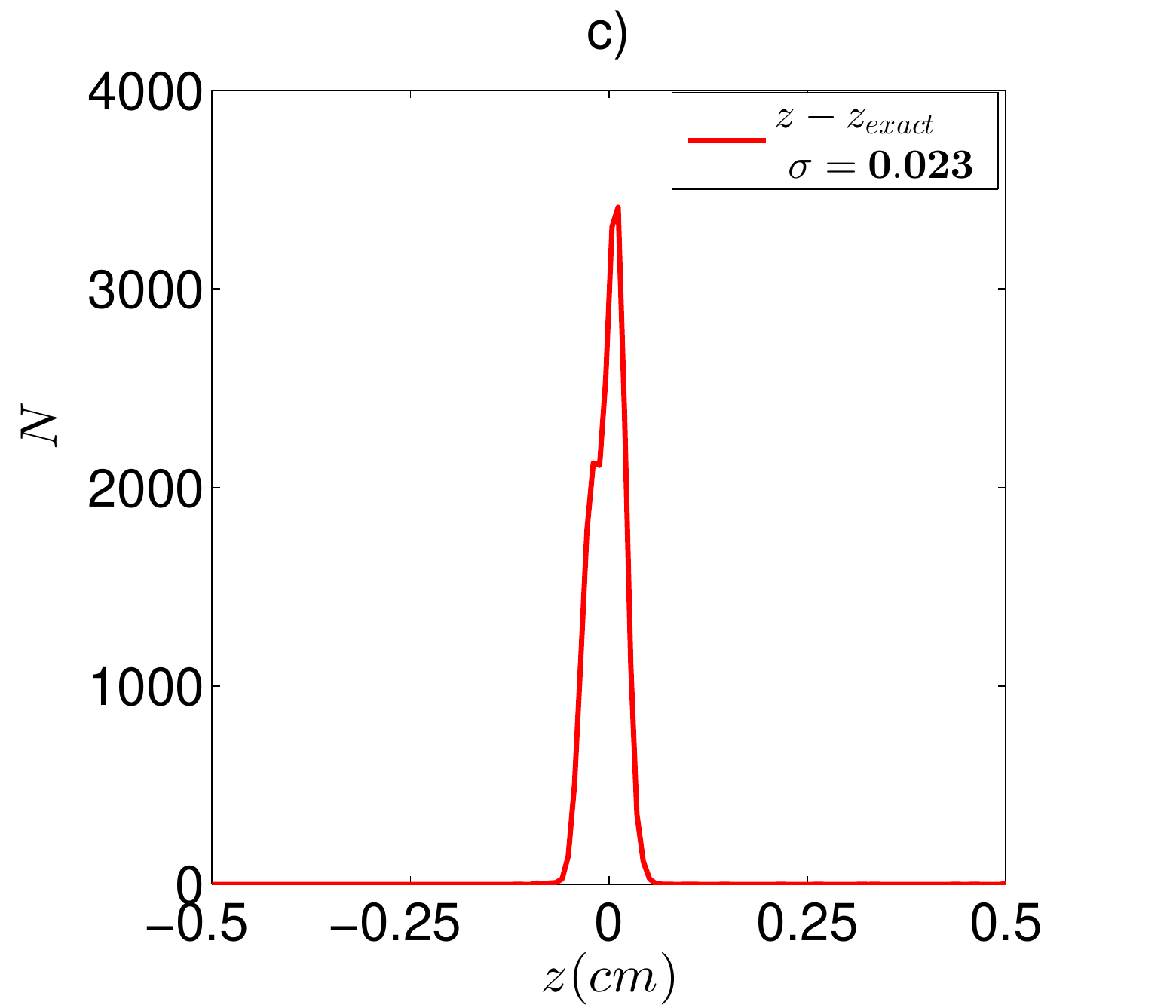}
\includegraphics[width=0.49\textwidth]{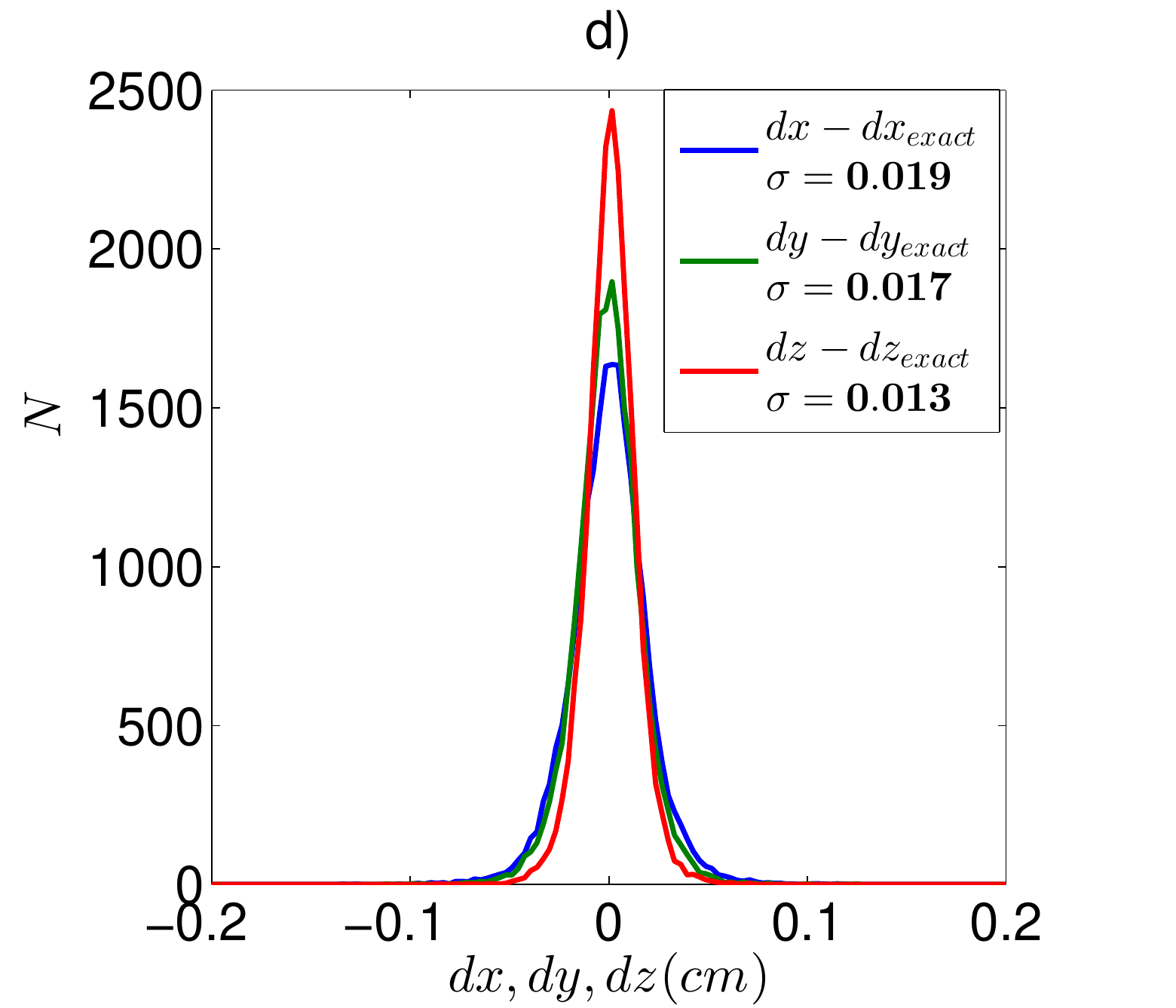}

\caption{a) White noise image used to create a pair of artificial images by geometric transform. b) 3D plot of an artificial wave deformation used to test the stereo-piv
reconstruction.  c) Distribution
of $z-z_{exact}$. We observe a dispersion $\sigma=\left\langle \left(z-z_{exact}\right)^{2}\right\rangle ^{1/2} =$0.023~cm.
d) Distributions of the displacements of a particle $(dx,dy,dz)$ (those have to be mutiplied by $dt^{-1}$=20 s$^{-1}$ to get the corresponding velocities in cm/s). \label{fig:fausse_image}}
\end{figure}

We present here tests of the accuracy of the method in the conditions of our experiment. The first test consists in generating numerically pairs of synthetic images of a deformed interface. This is done by creating a white noise random pattern in the physical $(x,y)$ coordinates (Fig. \ref{fig:fausse_image} a),  and an artificial sinusoidal deformation $z_{exact}$ with a typical amplitude that we encounter in our experiment (Fig. \ref{fig:fausse_image} b). We create artificial images  for each camera by the transform functions Eq. \ref{eq:geometric_transform}. We then apply our algorithm to reconstruct the wave field. The potential errors coming from the calibration are thus removed and the only remaining errors come from the image correlation maximization and from the stereoscopic reconstruction algorithm. The difference between the reconstructed height $z$, obtained with the cameras $1$ and $2$, and the artificial deformation $z_{exact}$ is plotted in Fig. \ref{fig:fausse_image} c. 

To quantify the error, we compute the root mean square $\sigma_{z-z_{exact}}=\left\langle \left(z-z_{exact}\right)^{2}\right\rangle ^{1/2}=0.023$~cm, where$\left\langle \right\rangle $ represent the mean over pixels. This estimated error of $\sigma_{z-z_{exact}}=0.023$~cm corresponds roughly to $0.13$ pix, which is typical for correlation error (see \cite{Scarano1999}).
Concerning the displacement of a particle$(dx,dy,dz)$ given by the PIV stereo in Fig. \ref{fig:fausse_image} d, we observe an error of the same order of magnitude for $dz$ and a slightly smaller one for the horizontal components. 

\begin{figure}
\includegraphics[width=0.49\textwidth]{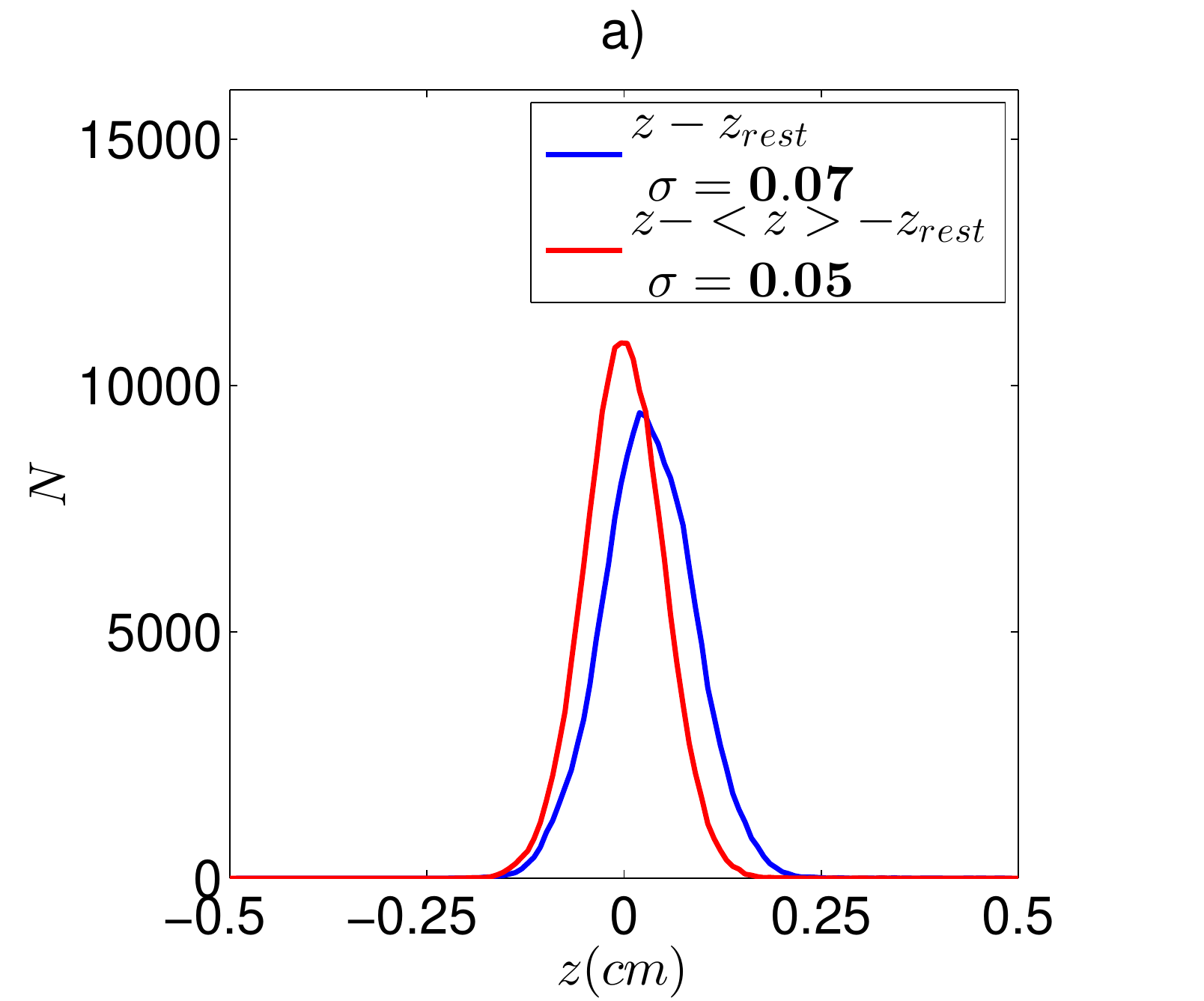}
\includegraphics[width=0.49\textwidth]{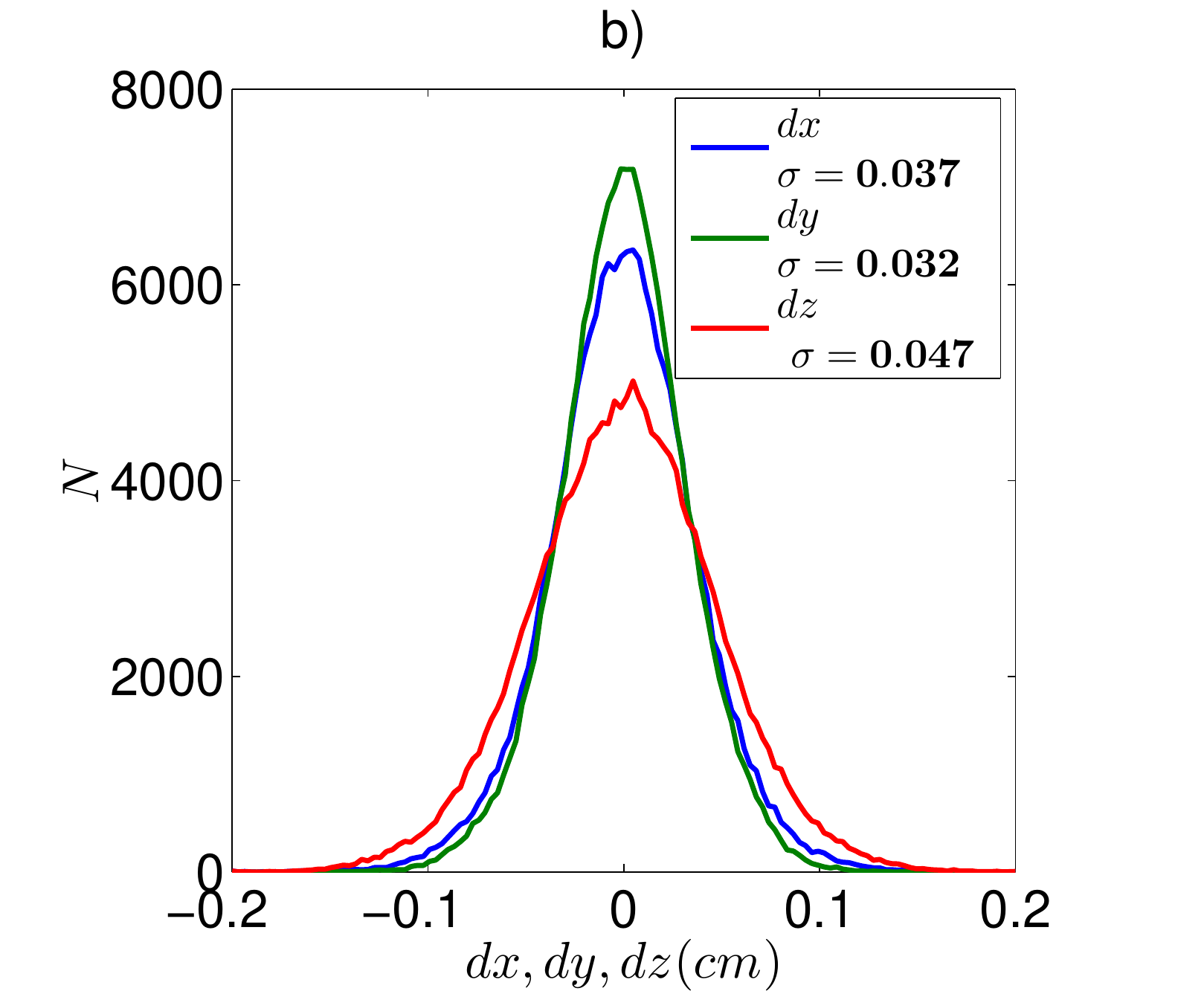}

\caption{a) Distribution of $z-z_{rest}$ (blue) and $z-<z>-z_{rest}$ (red). The dispersions are respectively $\sigma=0.05$~cm and $\sigma=0.07$~cm which is close to our particles diameter ($0.07$~cm). b) Distribution of the three displacements $dx,dy$ and $dz$ given by the PIV-stereo for the stationary surface (displacements must be zero). Those have to be multiplied by $dt^{-1}$=20 to get the corresponding velocity errors. 
\label{fig:rest}}
\end{figure}

The error induced by the calibration can come in several ways: wrong approximations of the geometric calibration model, imperfection of the grid used for calibration, post-calibration camera tilt. To quantify these static errors we consider the reconstitution of the water surface at rest, where $z$ should be perfectly flat. Since displacements are weak, we use directly the most separated camera pair $(1-2)$. Fig. \ref{fig:rest}a shows the difference between the mean level measured with the capacity probes $z_{rest}$ and $z$ obtained from the stereoscopic reconstitution. The blue curve shows the distribution of $z-z_{rest}$ integrated over $10$ images (this is sufficient to reach convergence since a single field already contains more than 10~000 measurement points). We observe a r.m.s. $\sigma=0.07$~cm which corresponds roughly to $0.4$ pix. We also observe that the mean is not quite zero, indicating a systematic error in the measurement. As the error remains small, we may subtract the temporal mean of each pixel. This operation is displayed with the red curve and shows a reduced dispersion of  $\sigma=0.05$~cm or $0.3$ pix.

Fig. \ref{fig:rest} b shows the three displacements $dx$, $dy$ and $dz$ obtained with the stereo-PIV, which should be equal to zero in this static test. The dispersion for the vertical velocity $dz$ is equivalent to that of the stereo measurement. We observe a slightly better accuracy for the two other components.

\begin{figure}
\includegraphics[width=0.49\textwidth]{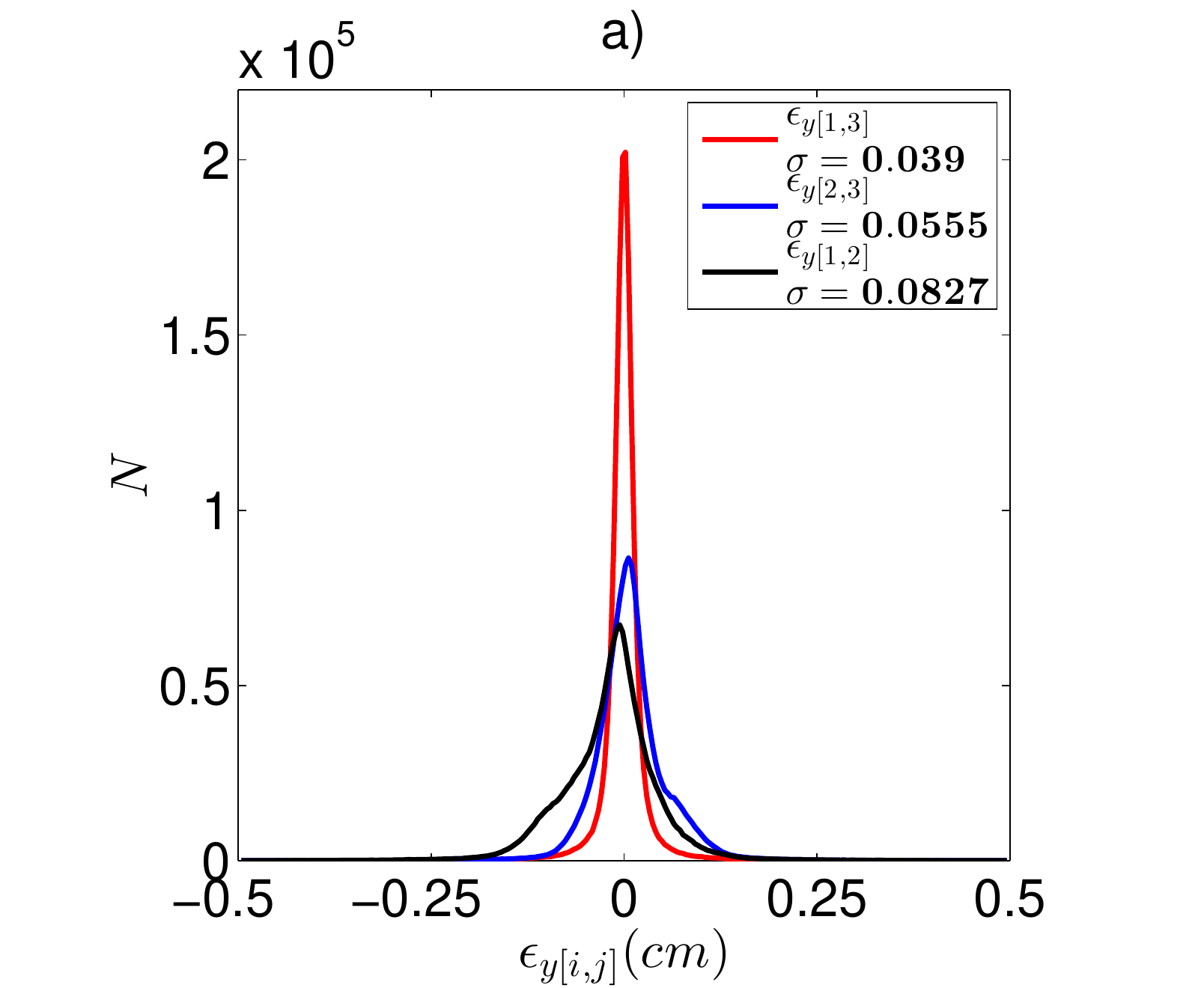}
\includegraphics[width=0.49\textwidth]{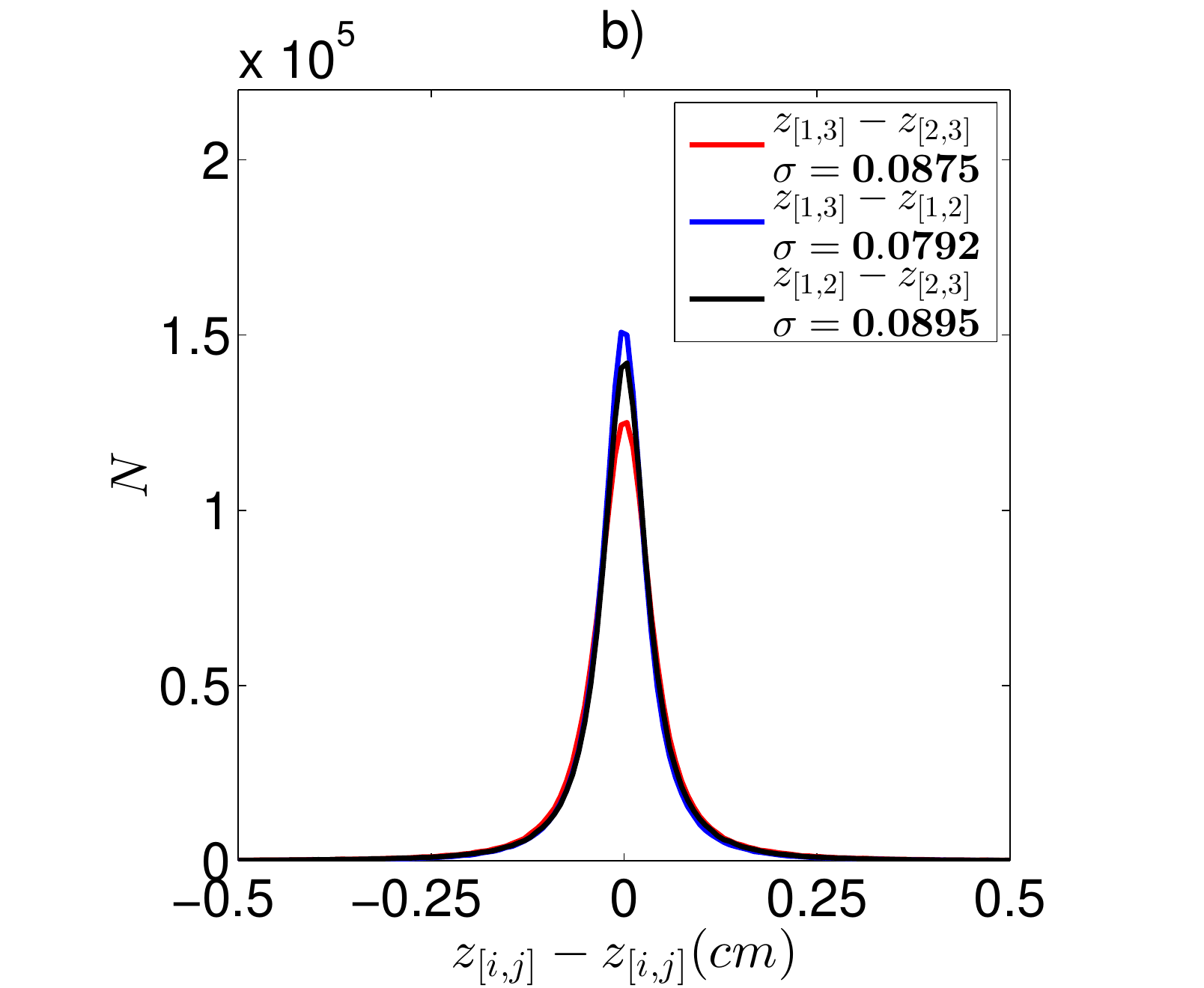}
\caption{a) pdf of the error $\epsilon_{Y}$ given by Eq. \ref{eq:error_X}. Using Eq. \ref{eq:Z_stereo}, we can deduce the r.m.s errors in $z$, $\sigma_{z(1-3)}=0.050$~cm, $\sigma_{z(2-3)}=0.058$~cm and $\sigma_{z(2-3)}=0.048$~cm.
b) pdf of the differences between the three measurements of $z$ given by the three cameras pairs $(1-3)$, $(2-3)$ and $(1-2)$ . 
\label{fig:diff_H}} 
\end{figure}

The previous tests allow us to evaluate the accuracy for an ideal stationary situation. However, the precision for real dynamical fields may be
less good. A first estimate can be made by the internal consistency of the stereoscopic reconstruction, using the error $\epsilon_Y$ determined by  Eq. \ref{eq:error_X}.  Fig. \ref{fig:diff_H} a) displays the pdf of this error for the three camera pairs. We observe non-Gaussian distribution with r.m.s ranging from $\sigma=0.04$~cm to $\sigma=0.08$~cm. With an hypothesis of isotropy on the error, we can assume that the unknown error $\epsilon_X$  has the same rms $\sigma$. Then from Eq. \ref{eq:Z_stereo}, we can estimate the rms $\sigma_z$ of the error in $z$ as  $\sigma/[(D_{xb}-D_{xa})^2+(D_{yb}-D_{ya})^2]^{1/2}$. This yields $\sigma_{z(1-3)}$=0.050~cm, $\sigma_{z(2-3)}$=0.058~cm and $\sigma_{z(1-2)}$=0.048~cm.

Thanks to our multiple cameras, we have in addition the possibility
to compare the three independents measurements of the wave displacement. Fig. \ref{fig:diff_H} b shows the differences between the three surfaces measurements obtained by the cameras pairs $(1-3)$, $(2-3)$ and $(1-2)$. We observe a quasi-equal distribution with roughly $\sigma=0.08$~cm. Since constant errors due to calibration are removed by the subtraction of the temporal mean field, we can assume that the errors from both pairs become uncorrelated, so the error on the difference squared is just the sum of the error squared for each measurement. Thus the r.m.s for each measurement is the r.m.s on the difference divided by $\sqrt{2}$, leading to $\sigma_{z} = 0.057$~cm. This  is consistent with the previous estimation of $\sigma_{z}$ obtained from  Eq. \ref{eq:error_X}. Comparing with the distribution of waves (Fig. \ref{fig: PDFHUVW} a)), the error represent about $1.5\%$ of the maximum amplitude, corresponding also to $0.3$ pix. Comparing with the previous test on the stationary surface, we therefore conclude the stereo measurement performs similarly in dynamical conditions. 

\begin{figure}
\includegraphics[height=7cm]{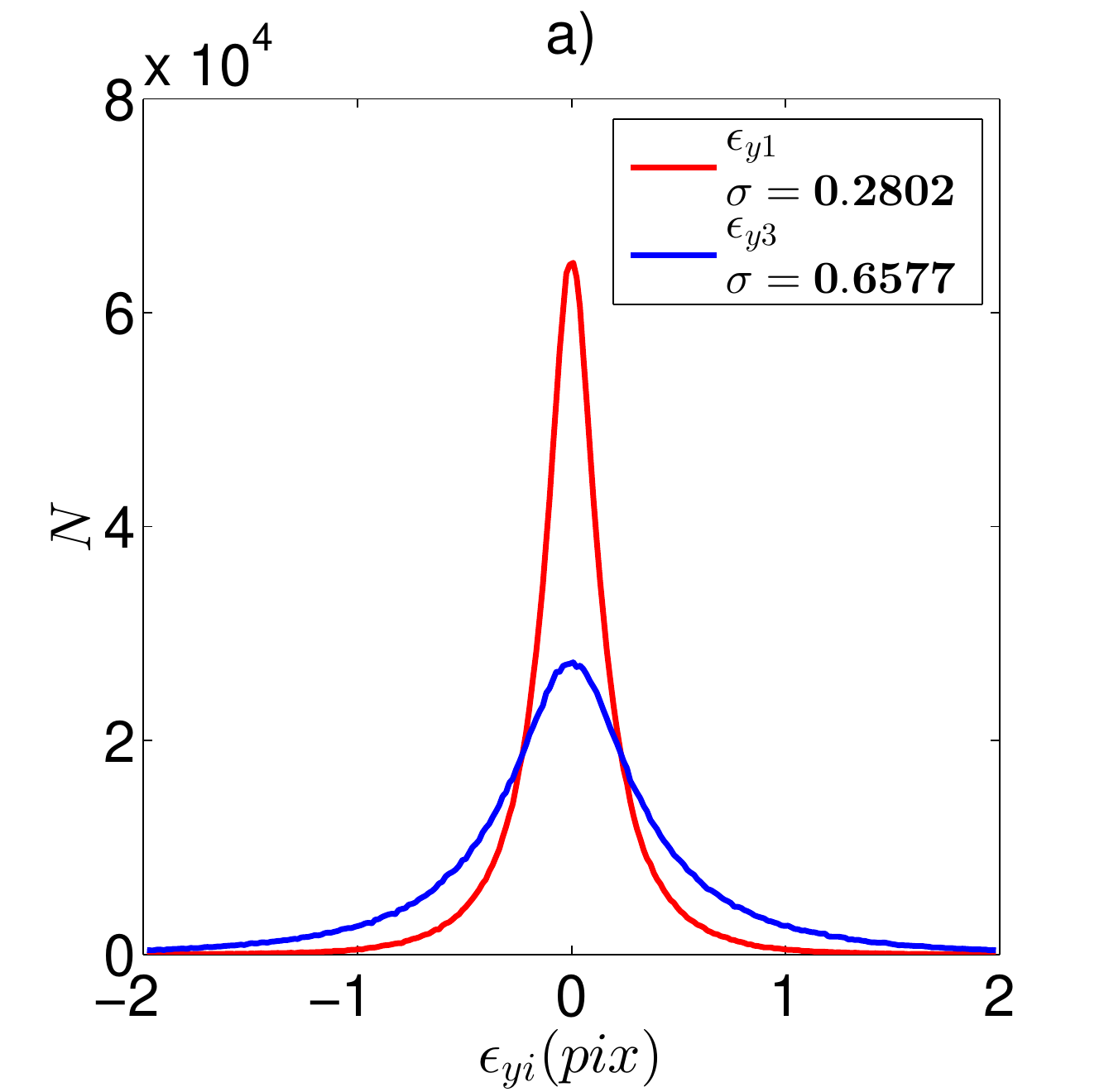}
\includegraphics[height=7cm]{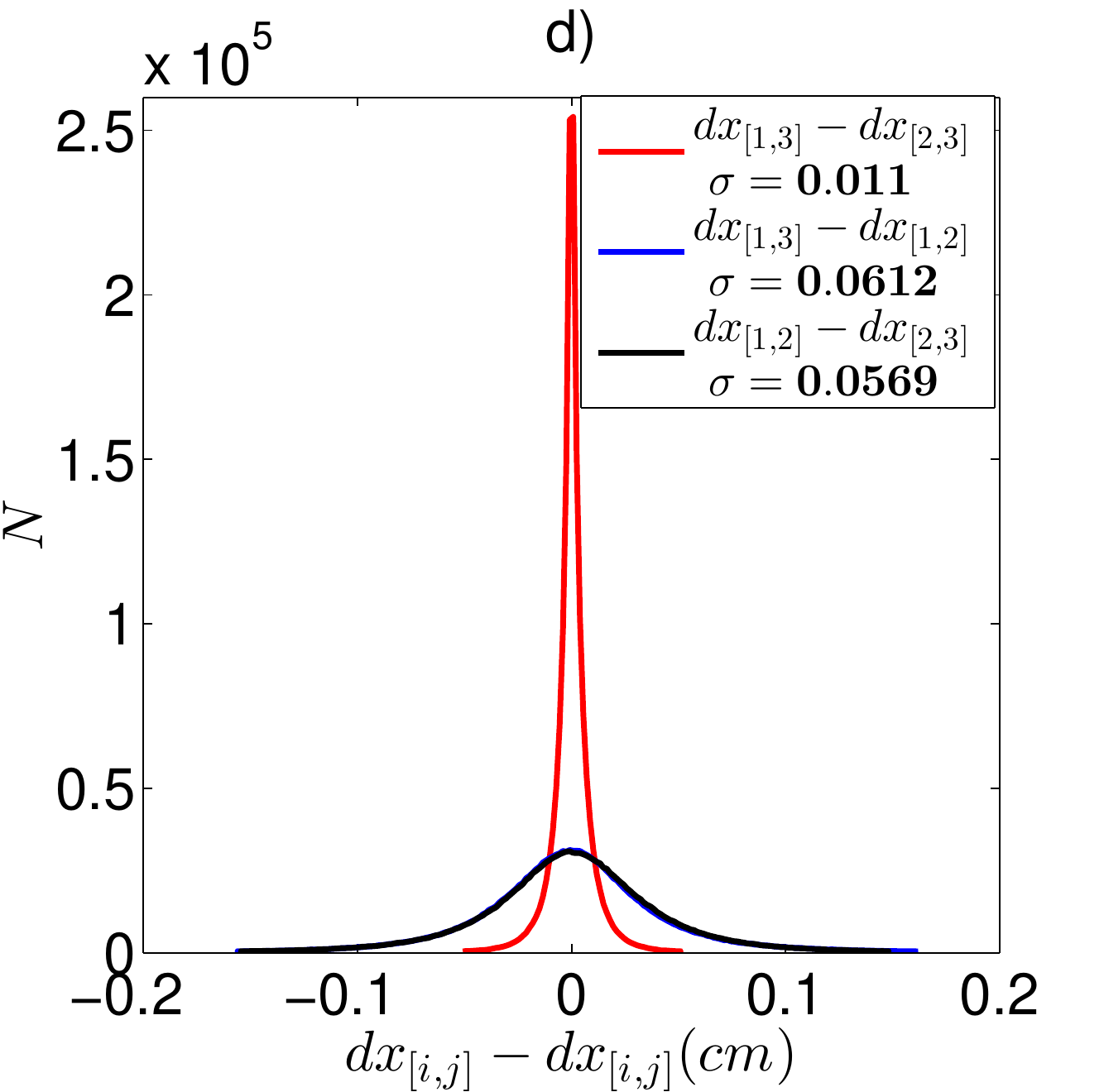}
\includegraphics[height=7cm]{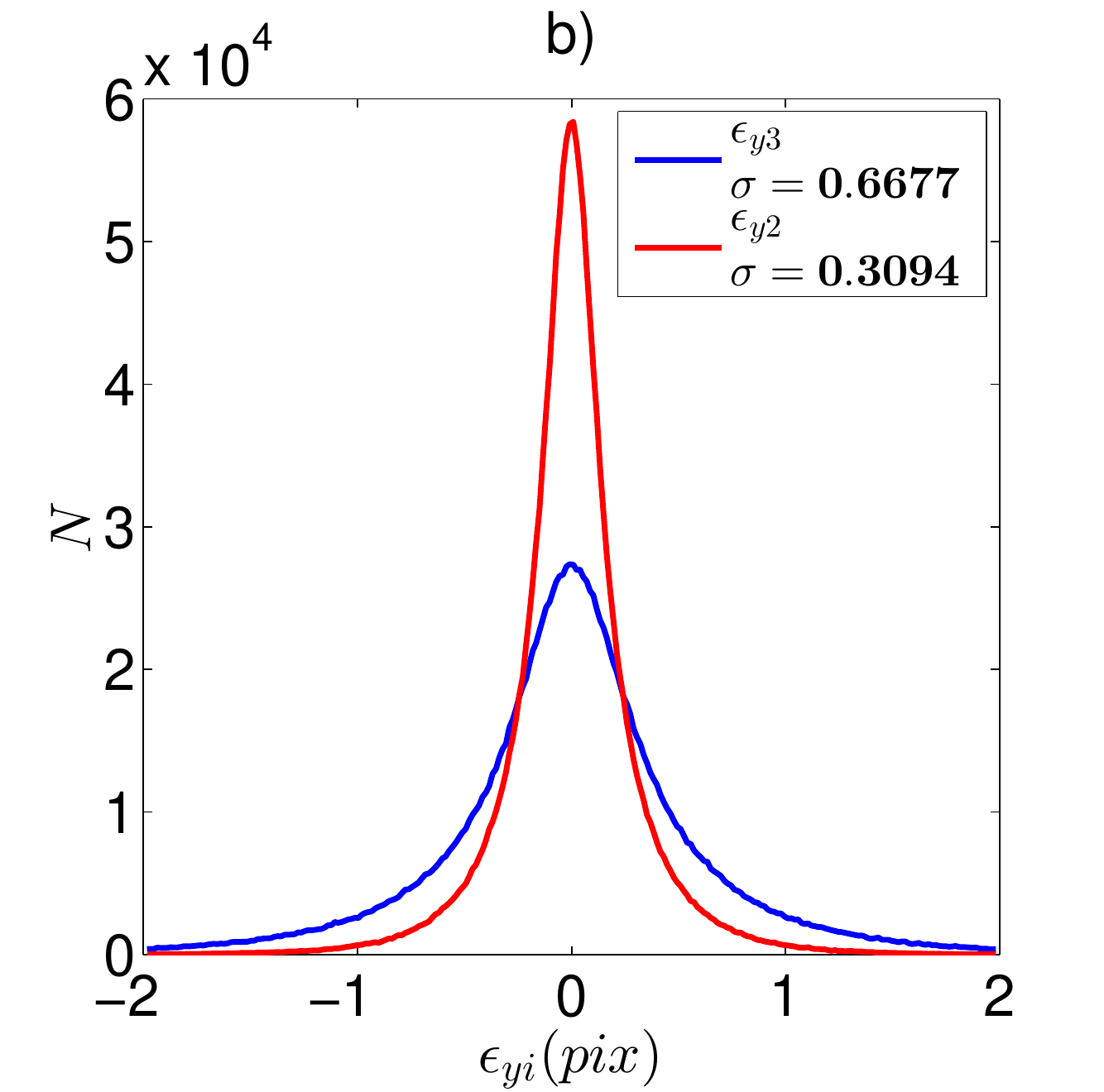}
\includegraphics[height=7cm]{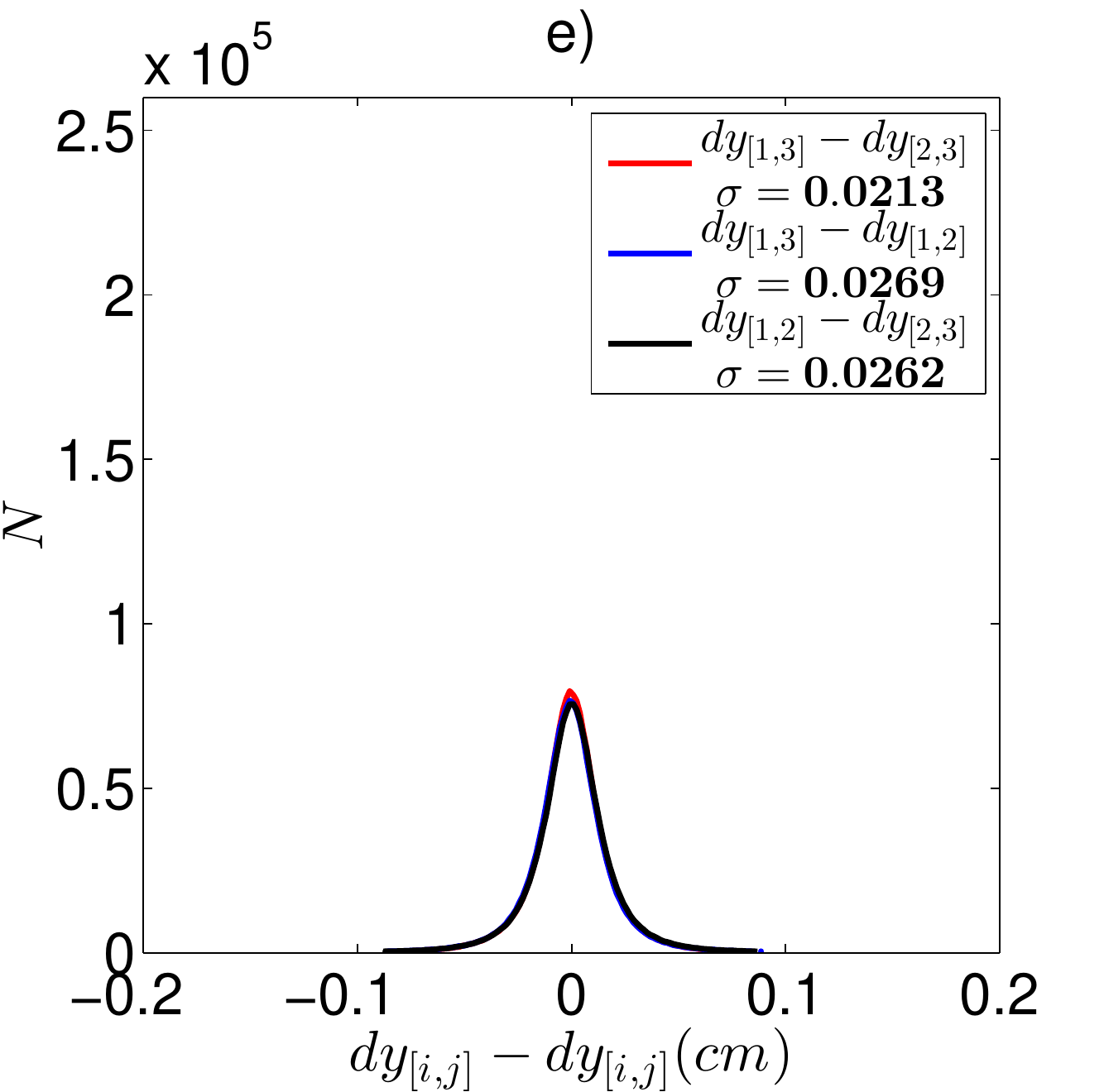}
\includegraphics[height=7cm]{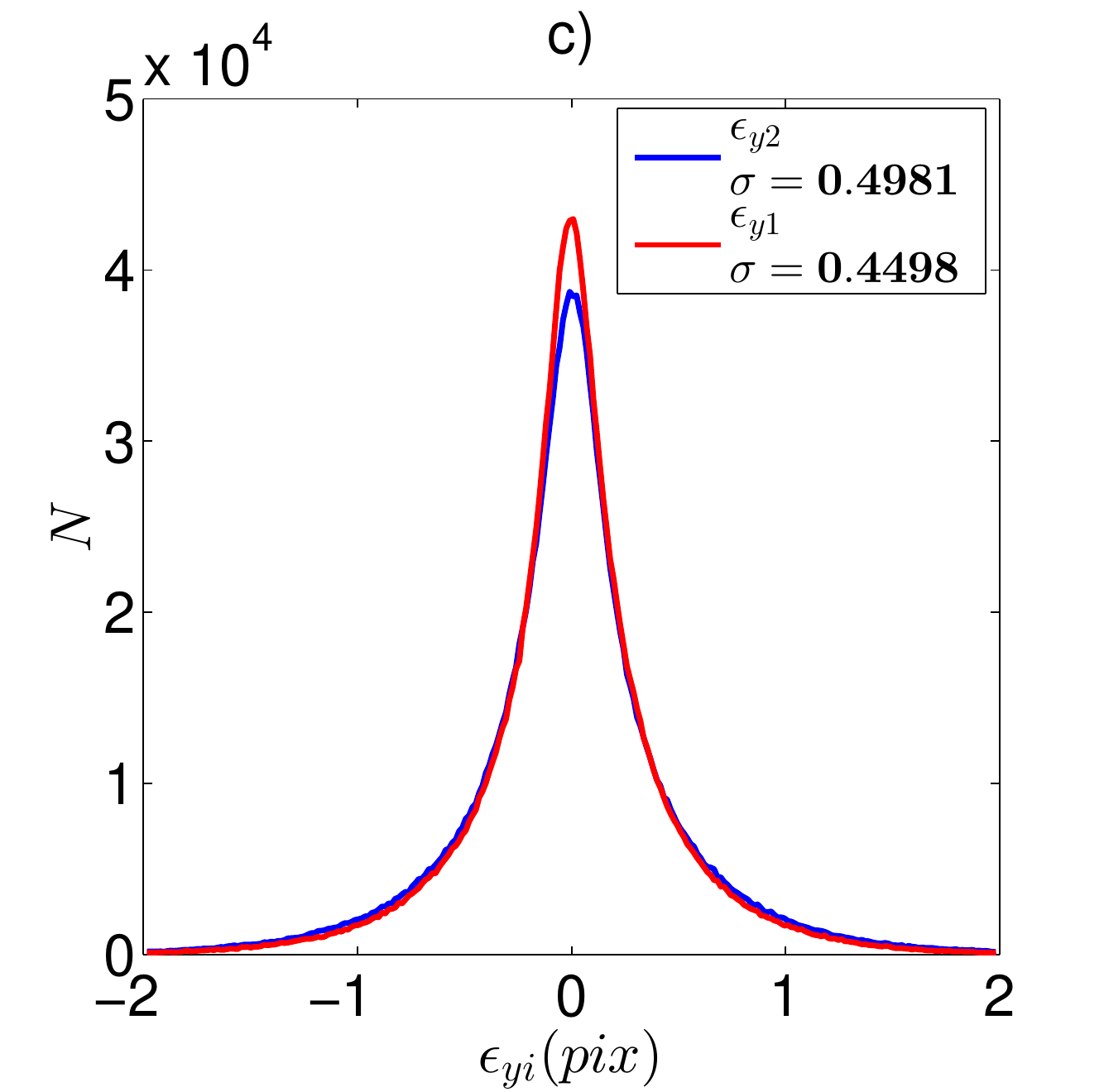}
\includegraphics[height=7cm]{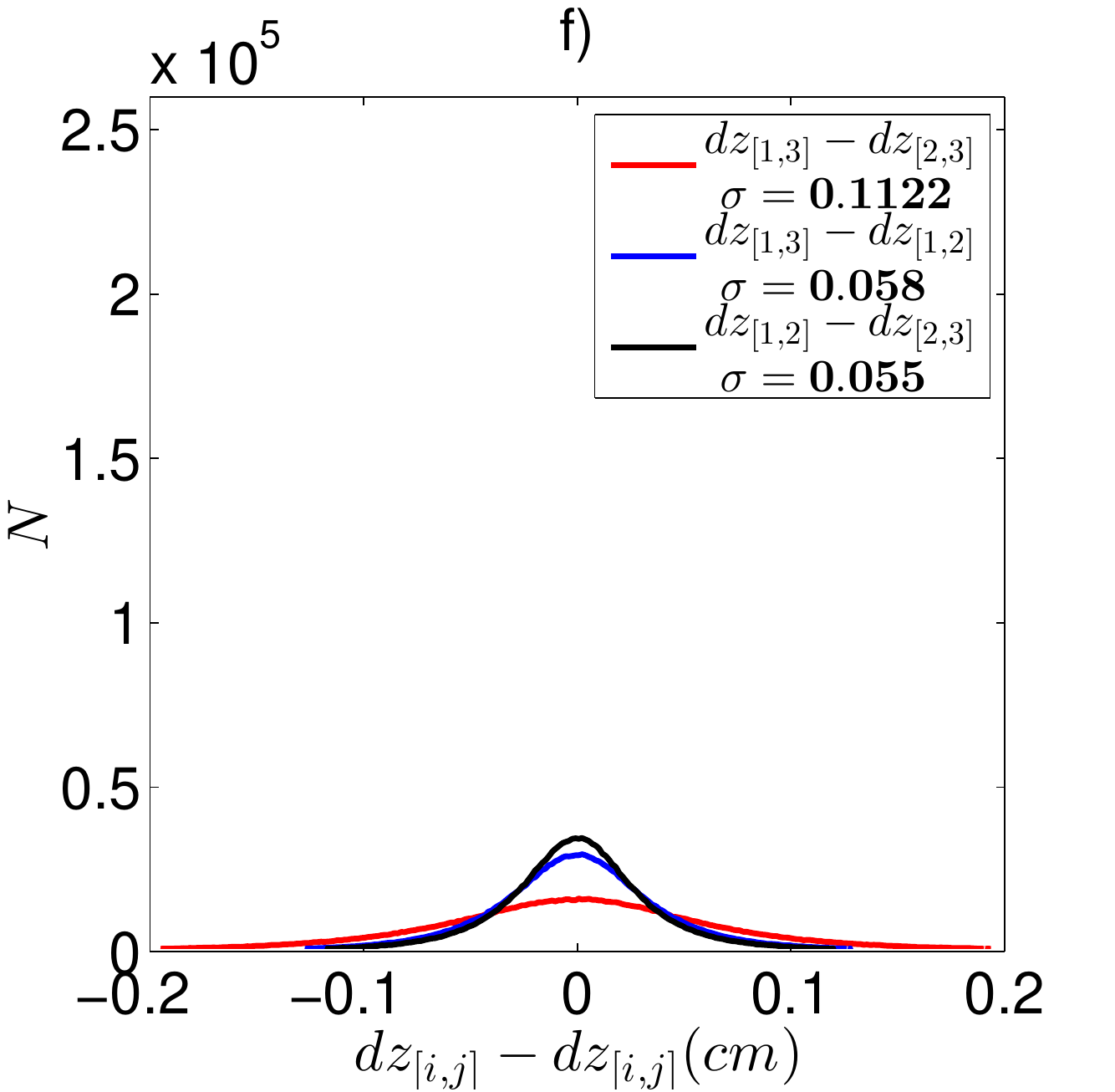}

\caption{a)b)c) Error $\epsilon_{yi}$ given by Eq. \ref{eq:epsPIV}. $\epsilon_{xi}$ is
negligible in our configuration. Using Eq. \ref{eq:error_PIV}, we can
estimate the global error : $\epsilon'_{[1,3]}=0.36$ pix $\epsilon'_{[1,3]}=0.37$ pix
$\epsilon'_{[1,2]}=0.34$ pix d) Differences of the three estimations
of the vertical displacement $dz$.e) Differences of the three estimations
of the horizontal displacement $dx$.
\label{fig:a)Comp_Vitesse}}
\end{figure}

We now perform the same analysis on the PIV measurement. Fig. \ref{fig:a)Comp_Vitesse}a,b,c) shows $\epsilon_{yi}$ described by Eq. \ref{eq:epsPIV} for our three camera pairs. The index $i$ denotes the camera number. Like for the stereo, $\epsilon_{xi}$ is negligible compare to $\epsilon_{yi}$ due to the positioning of the cameras along the $x$ axis. From these values, we can compute the global error estimate $\epsilon'$ given in Eq. \ref{eq:error_PIV}.

We observe a global error equivalent for the three pairs : $\epsilon'_{(1-3)}=0.36$ pix, $\epsilon'_{(1-3)}=0.37$ pix and $\epsilon'_{(1-2)}=0.34$ pix. However, the accuracy is not identical on each components of the displacement. We can compare the different estimations given by the three cameras pairs. This is displayed in Fig. \ref{fig:a)Comp_Vitesse}~d,e,f) for $dx$, $dy$ and $dz$. For $dz$ we observe that the worst estimation is when the camera $3$ is present in both pairs (red curve in f). This is explained by the lack of sensitivity on the vertical displacement for this camera which is normal to the field and has only a direct measurement of $dx$ and $dy$.
Thus, inversely, the best accuracy on $dx$ (red curve in d) is reached when the camera $3$ is used in both pairs. Errors on $dy$ remains equal, due to the common angle of sight for the three cameras. Lowest errors are roughly $\epsilon_{dz}\sqrt{2}=0.040$, which is consistent with the previous estimation in the state of rest.

\begin{figure}
\includegraphics[height=7cm]{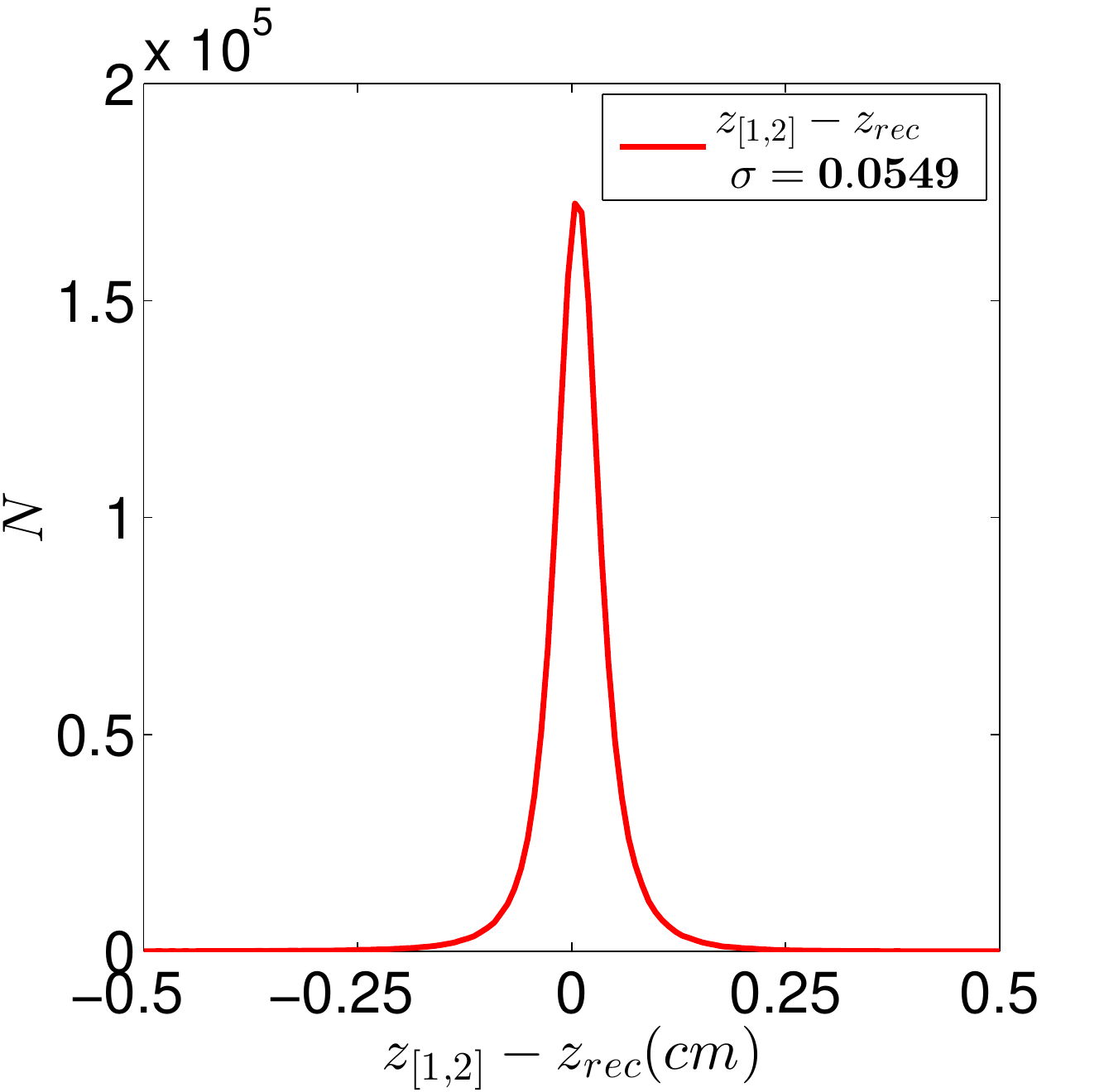}
\caption{Difference between the recombination of two low-angle cameras: $z_{rec}$
with the direct large angle measurement $[1,2]$ .\label{fig:Rec}}
\end{figure}

A wide angle between the cameras is favorable for the precision of the stereoscopic measurement, but the image matching may become difficult to achieve. The introduction of an intermediate camera is then useful to combine two
stereo measurements with  lower angles. Fig. \ref{fig:Rec} shows the distribution of the difference between the direct measurement $z_{(1-2)}$ and the
recombination with the two low-angle cameras pairs $z_{rec}$. We observe an error near $\sigma=0.05$~cm, which is comparable to the previous estimated errors for the stereoscopic mapping. Thus, this technique may be used to increase the vertical sensitivity of $dz$. 

\begin{figure}
\includegraphics[width=8cm]{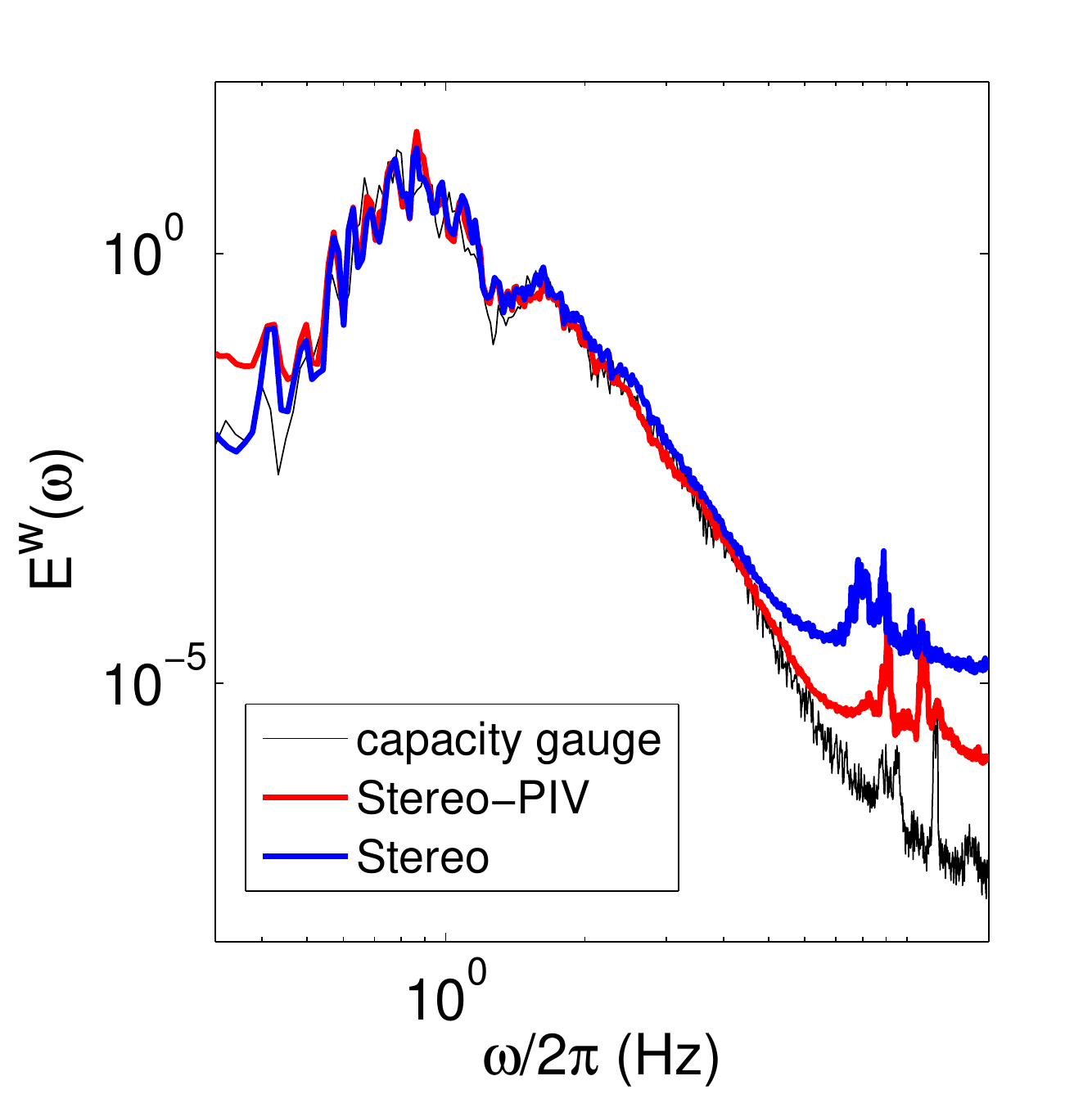}

\caption{Temporal spectrum $E^{z}(\omega)=\left\langle \left|z(\omega)\right|^{2}\right\rangle $
. Black line: capacity probes. Blue line: Stereo. Red line : Stereo-PIV\label{fig:Temporal-spectrum-}}
\end{figure}

As a final confirmation of the global method, we can compare the temporal wave power spectrum $E^{z}(\omega)=\left\langle \left|\hat{z}(\omega)\right|^{2}\right\rangle $ with the one obtained from the local probes. Capacitive probes are assumed to have
a better dynamical range of measurement and so they are a good indication for the temporal quality of the stereo-piv measurements.
 Fig. \ref{fig:Temporal-spectrum-} displays the measurement obtained from the three methods. The black solid line is the measurement done with
capacitive probes and the blue is the Stereo. We observe a good agreement up to 5~Hz, which represents 4 decades. The red curve is the integration of the vertical velocity $w$. As seen before, the link between $w$ and $d\eta/dt$ involves the non-linear term $-\mathbf{u}\nabla\eta$. However, from Fig. \ref{fig:nonlin} it remains negligible in our case making the integration of $w$
valid. We observe a gain of about one decade on the dynamics. This point emphasizes the input of the PIV to increase the accuracy of the measurements of the wave displacement. The common peak of noise visible near 10~Hz comes from the vibration of the structure. The small differences between the power spectrum of the probes and the ones of the 3D reconstruction may arise from the different localizations of the measures. Although the system is well homogeneous, some minor differences are still present. 

\section{Conclusion}
The stereo-PIV allows us to measure a velocity field on a fluctuating surface. The deformation of the interface as well as the velocity field are obtained with sub-pixel vertical accuracy of about $0.3$ pix in the best configuration. Here the measurement area is about $2\times1.5$~m$^2$ but wider fields of view would be accessible without loss on the spatial resolution. When the deformation of the surface is too important to allow the computation of the cross-correlations for the stereo, an additional camera is used as an intermediate step. 

 The tests presented in this paper validate the technical relevance of this method for measurement of surface waves. The combination of the two methods allows a reconstruction with a good sensitivity at all scales. The low frequency deformations are well evaluated with the stereoscopic reconstruction while the fast dynamical field are measured with PIV. This property leads to increase significantly the quality of the wave spectra. In the case of our surface gravity waves measurement, we observe a gain of about 1 order in magnitude on the power spectrum for the higher frequencies compared to the stereoscopic reconstruction alone. Furthermore the horizontal velocity components are interesting by themselves, giving for instance a direct mapping of the nonlinear term $\mathbf{u}.\nabla \eta$.

In this experiment the spatial resolution is mainly limited by the floating particles. The final correlation box is roughly $30\times30$ pixels and it is difficult to go below because of non-homogeneity of the particle seeding. This is principally caused by the natural attraction of two object floating on the surface \cite{Kralchevsky2000}. The deformation caused by the capillarity induces a depression of the free surface between the two particles and tends to attract them over a distance of ten diameters of particles \cite{Gifford1971} . The reduction of this effect can be obtained by decreasing the diameter of the particles. But then particles are more prone to be entrained by turbulence below the surface. Fortunately, if the waves are sufficiently strong, the induced mixing tends to break the clusters of particles. 

Although this method is well adapted for surface waves measurements, there are other possibilities of applications where the knowledge of the velocity is useful. In particular, this permits to measure the surface vorticity that may be interesting for surface flow studies or wave-current interactions for instance.

\begin{acknowledgements}
%\section{aknowledgements}
This project has received funding from the European Research Council (ERC) under the European Union's Horizon 2020 research and innovation programme (grant agreement No 647018-WATU). The  software has been developed within the JRA `COMPLEX' and `RECIPE' of the European Integrated Infrastructure Initiative Hydralab+.  NM is supported by Institut Universitaire de France. 
\end{acknowledgements}

\section{Appendix: practical implementation}
\label{Appendix}

All the programs described below are freely available with graphic interface in the Matlab toolbox UVMAT http://servforge.legi.grenoble-inp.fr/projects/soft-uvmat . 

\subsection{Calibration grid}

\begin{figure} 
\includegraphics[width=0.49\textwidth]{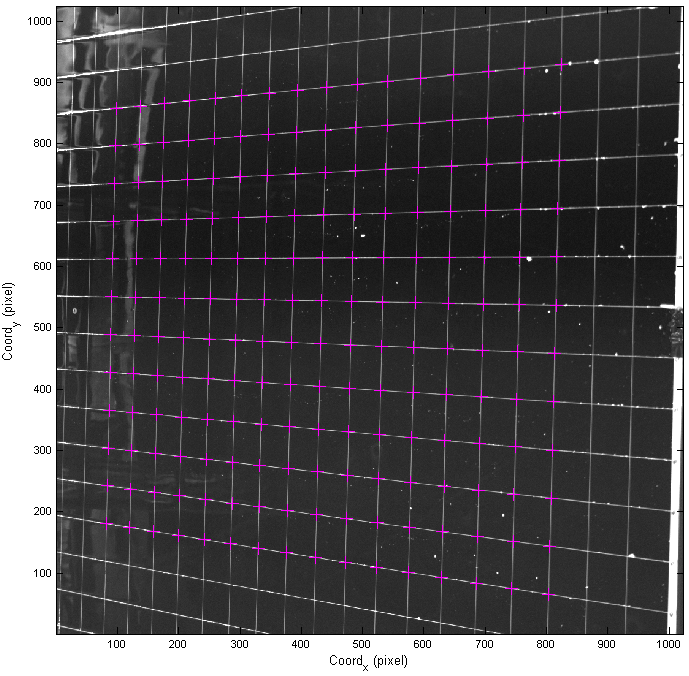}
\includegraphics[height=0.49\textwidth]{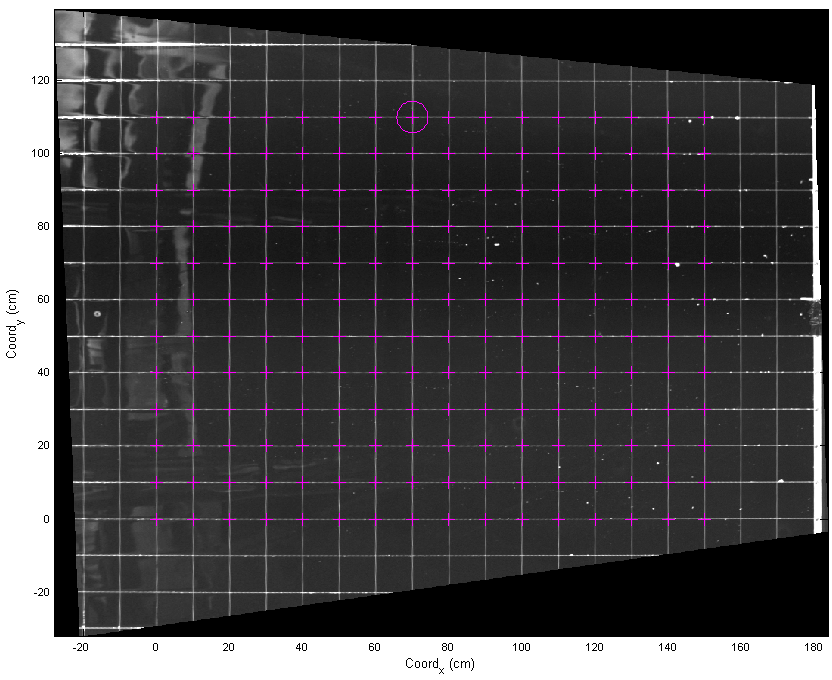}
\caption{image of the calibration grid with the detected reference points, a) in image coordinates, b) in physical coordinates at $z=0$\label{fig:grid_calib}}
\end{figure}

The first step in camera calibration is to take image of a calibration
grid with well identified points that evenly cover the field of view,
with different distances to the camera to provide 3D information.
For that purpose we use a light aluminium  frame $2.2\times2.2$~m$^2$ which
holds two perpendicular arrays of stretched white strings with mesh
$10$~cm, providing a plane grid of $20\times20$ reference points with precision
$1$~mm. An image of this grid, as seen from one of the cameras, is shown in Fig. \ref{fig:grid_calib}. The image also shows the points 
which are automatically detected at the crossing of the strings.This detection relies on a projection transform of the type Eq. \ref{eq:projection} to get a square grid image, followed by a detection of image maxima along lines and columns (using sub-pixel quadratic interpolation near the maximum).

To get the 3D calibration with the plane grid, we take images
of the grid with different angles of sight, and use the calibration
method of Tsai \cite{Tsai1987}, further improved by J. Heikkil\"a and O. Silv\'en \cite{Heikkila1997}. We use a Matlab toolbox
implementation `Camera Calibration Toolbox` provided by Jean-Yves Bouguet which is the numerical application of the method proposed by Z. Zhang \cite{Zhang1999}). This method relies on the perspective
effects and does not require any measurement of the angle of sight:
the grid is just manually tilted with different orientations (in the
absence of water). Typically five tilted images are used, as shown in Fig. \ref{fig:grid_calib}, involving about 200 calibration points each, well spread over the main part of the image. We then put the grid in horizontal position to
get the reference plane $z=0$ at the height expected for the water
surface. 

\subsection{The pinhole camera model}

%\begin{figure} 
%\includegraphics[height=7cm]{Pinhole_sketch.png}
%\caption{sketch of the pinhole camera model\label{fig:pinhole_sketch}}
%\end{figure}

The calibration method relies on the classical pinhole camera model.
The transform from the physical coordinates $(x,y,z)$ to the image
coordinates is performed by the following steps: 
\begin{enumerate}
\item A rotation and translation to express position in the 3D coordinates
$(x_{c},y_{c},z_{c})$ linked to the camera sensor, with origin at
the center of the optical axis on the image sensor, and $z_{c}$ along
the optical axis outward (see sketch in Fig. \ref{fig:pinhole_sketch}).
\begin{equation}
\left[\begin{array}{c}
x_{c}\\
y_{c}\\
z_{c}
\end{array}\right]=\left[\begin{array}{ccc}
r_{1} & r_{2} & r_{3}\\
r_{4} & r_{5} & r_{6}\\
r_{7} & r_{8} & r_{9}
\end{array}\right]\left[\begin{array}{c}
x\\
y\\
z
\end{array}\right]+\left[\begin{array}{c}
T_{x}\\
T_{y}\\
T_{z}
\end{array}\right]\label{eq:rotation}
\end{equation}

\item A projection on the sensor plane. 
\begin{equation}
\begin{array}{c}
X'=x_{c}/z_{c}\\
Y'=y_{c}/z_{c}
\end{array}\label{eq:projection}
\end{equation}
Those correspond to the tangent of the viewing angles.
\item A rescaling factor and nonlinear distortion to express the coordinates
$X,Y$ on the sensor in pixels. Our optical system provides a limited
distortion, well described by a quadratic correction in terms of the angular
distance to the optical axis $(X'^{2}+Y'^{2})^{1/2}$, 
\begin{equation}
\begin{array}{c}
X=f_{x}\,[1+k_{c}(X^{'2}+Y^{'2})]X'+C_{x}\\
Y=f_{y}\,[1+k_{c}(X'^{2}+Y'^{2})]Y'+C_{y}
\end{array}\label{eq:intrinsic}
\end{equation}
The 'focal length' is expressed in units of pixel size on the sensor,
so it could take a priori a different value $f_{x}$ and $f_{y}$ along each
axis for non-square pixels (but they are square in our cameras). For a focus at infinity, it should fit
with the true focal length of the objective lens (normalized by the
sensor pixel size), but slightly higher for a focus at close distance.
A geometric distortion has been introduced as a first order quadratic
correction assumed axisymmetric around the optical axis, with coefficient
$k_{c}$. The typical value obtained is $k_c\simeq -0.015$ which is a small quadratic deformation, with no need to higher order correction. The parameters $C_{x}$ and $C_{y}$ represents a translation
of the coordinate origin from the optical axis to the image lower
left corner, so it must be equal to the half of the pixel number in
each direction for a well centered sensor.
\end{enumerate}

The composition of the three transforms Eq. \ref{eq:rotation}, \ref{eq:projection} and \ref{eq:intrinsic} specifies for each camera the functions $X=F(x,y,z)\;,\;Y=G(x,y,z)$ introduced in section \ref{stereo}.
The transform is defined by 17 parameters, among which 5
are intrinsic $(f_{x},f_{y},k_{c},C_{x},C_{y})$, as they depend only
on the optical system, and the other ones are extrinsic, as they depend
on the rotation and translation of the camera with respect to its
environment. Note that the nine coefficients of the rotation matrix $r_{i}$ depend only on
3 independent parameters, which are the rotation angles, so there
are 6 independent extrinsic parameters among the twelves.

The calibration error can be estimated by applying the obtained transform $X=F(x,y,z)\;,\;Y=G(x,y,z)$ to the physical coordinates of each calibration point and compare them to their pixel coordinates in the camera sensor.
 Typically a maximum error of 1 pixel is obtained with a r.m.s. error of 0.3 pixel. This corresponds roughly to a precision of 1 mm in the determination of physical coordinates from the image.

\subsection{The reverse transform}

The equations Eq. \ref{eq:projection} can be expressed as the linear
system $x_{c}-Xz_{c}=0$, $y_{c}-Yz_{c}=0$, which writes, using Eq.\ref{eq:rotation}:
\begin{equation}
\begin{array}{c}
A_{11}x+A_{12}y+A_{13}z=X'T_{z}-T_{x}\\
A_{21}x+A_{22}y+A_{23}z=Y'T_{z}-T_{y}
\end{array}\label{eq:lin_sys}
\end{equation}
where

\begin{equation}
\begin{cases}
A_{11}=r_{1}-r_{7}X\,,\, & A_{12}=r_{2}-r_{8}X\:A_{13}=r_{3}-r_{9}X\\
A_{21}=r_{4}-r_{7}Y\,,\, & A_{22}=r_{5}-r_{8}Y\:A_{23}=r_{6}-r_{9}Y
\end{cases}\label{eq:coeff_A}
\end{equation}
If the points are in a known plane, providing a third linear relation
between $x,y,z$, this linear system Eq. \ref{eq:lin_sys} can be solved,
providing the physical coordinates from the angular image coordinates
$(X',Y')$. Restricting ourselves to the plane $z=0$, we get
a linear system of two equations for the unknown $(x,y)$, whose solution
is

\begin{equation}
\begin{cases}
x_{a}=\frac{-A{}_{22}(X'T_{z}-T_{x})+A{}_{12}(Y'T_{z}-T_{y})}{A{}_{11}A{}_{22}-A{}_{12}A{}_{21}}\\
y_{a}=\frac{-A{}_{21}(X'T_{z}-T_{x})+A{}_{11}(Y'T_{z}-T_{y})}{A{}_{11}A{}_{22}-A{}_{12}A{}_{21}}
\end{cases}\label{eq:image2phys}
\end{equation}

The angular image coordinates $X'$ and $Y'$ are obtained from the
image coordinates $X,Y$ of the sensor by solving the system of equations equation 
Eq. \ref{eq:intrinsic} which depends only on the intrinsic parameters.
Since the quadratic deformation is weak, it can be first inverted
linearly as 
\begin{equation}
\begin{cases}
X'\simeq(X-C_{x})f_{x}^{-1}\\
Y'\simeq(Y-C_{y})f_{y}^{-1}
\end{cases}
\end{equation}
Then in a second step, using these values of $X$ and $Y$ to estimate
the quadratic correction, 
\begin{equation}
\begin{cases}
X'=(X-C_{x})f_{x}^{-1}\:[1+k_{c}f_{x}^{-2}(X-C_{x})^{2}+k_{c}f_{y}^{-2}(Y-C_{y})^{2}]^{-1}\\
Y'=(Y-C_{y})f_{y}^{-1}\:[1+k_{c}f_{x}^{-2}(X-C_{x})^{2}+k_{c}f_{y}^{-2}(Y-C_{y})^{2}]^{-1}
\end{cases}\label{eq:intrinsic_reverse}
\end{equation}
This approximation is excellent as the quadratic correction is less than 1\% (as $k_c\simeq 0.02$ and $X'<1$), so the next order in the expansion would be of the order of 10$^{-4}$.

Combined with Eq. \ref{eq:image2phys}, this provides the explicit expression
of the physical coordinates $x_{a},y_{a}$ versus the image coordinates
$(X,Y)$.

\subsection{Jacobian matrix\label{sub:Jacobian-matrix}}

Combining the differential of Eq. \ref{eq:projection}
\begin{equation}
\begin{array}{c}
dx_{c}=X'dz_{c}+z_{c}dX'\\
dy_{c}=Y'dz_{c}+z_{c}dY'
\end{array}\label{eq:projection-1}
\end{equation}
and the differential of Eq. \ref{eq:rotation}, 
\begin{equation}
\left[\begin{array}{c}
dx_{c}\\
dy_{c}\\
dz_{c}
\end{array}\right]=\left[\begin{array}{ccc}
r_{1} & r_{2} & r_{3}\\
r_{4} & r_{5} & r_{6}\\
r_{7} & r_{8} & r_{9}
\end{array}\right]\left[\begin{array}{c}
dx\\
dy\\
dz
\end{array}\right]\label{eq:rotation-1}
\end{equation}
yields
\begin{equation}
\begin{array}{c}
(r_{1}-r_{7}X')dx+(r_{2}-r_{8}X')dy+(r_{3}-r_{9}X')dz=(r_{7}x+r_{8}y+r_{9}z+T_{z})dX'\\
(r_{4}-r_{7}Y')dx+(r_{5}-r_{8}Y')dy+(r_{6}-r_{9}Y')dz=(r_{7}x+r_{8}y+r_{9}z+T_{z})dY'
\end{array}\label{eq:projection-1-1}
\end{equation}
From which the Jacobian matrix can be calculated,
\begin{equation}
\left[\begin{array}{ccc}
\frac{\partial X'}{\partial x} & \frac{\partial X'}{\partial y} & \frac{\partial X'}{\partial z}\\
\frac{\partial Y'}{\partial x} & \frac{\partial Y'}{\partial y} & \frac{\partial Y'}{\partial z}
\end{array}\right]=(r_{7}x+r_{8}y+r_{9}z+T_{z})^{-1}\left[\begin{array}{ccc}
r_{1}-r_{7}X' & \:r_{2}-r_{8}X' & \:r_{3}-r_{9}X'\\
r_{4}-r_{7}Y' & \:r_{5}-r_{8}Y' & \:r_{6}-r_{9}Y'
\end{array}\right]
\end{equation}
 This has to be combined with the quadratic transform, although it
is a small correction in our experiments. By the chain rule for the
two transforms, we have 
\begin{equation}
\left[\begin{array}{ccc}
\frac{\partial X}{\partial x} & \frac{\partial X}{\partial y} & \frac{\partial X}{\partial z}\\
\frac{\partial Y}{\partial x} & \frac{\partial Y}{\partial y} & \frac{\partial Y}{\partial z}
\end{array}\right]=\left[\begin{array}{cc}
\frac{\partial X}{\partial X'} & \frac{\partial X}{\partial Y'}\\
\frac{\partial Y}{\partial X'} & \frac{\partial Y}{\partial Y'}
\end{array}\right]\times\left[\begin{array}{ccc}
\frac{\partial X'}{\partial x} & \frac{\partial X'}{\partial y} & \frac{\partial X'}{\partial z}\\
\frac{\partial Y'}{\partial x} & \frac{\partial Y'}{\partial y} & \frac{\partial Y'}{\partial z}
\end{array}\right]
\end{equation}

The differentiation of Eq. \ref{eq:intrinsic} yields

\begin{equation}
\begin{array}{c}
f_{x}^{-1}dX=[1+3k_{c}X'^{2}+Y'^{2}]\:dX'+2X'Y^{'}\:dY'\\
f_{y}^{-1}dY=2X'Y^{'}\:dX'+[1+3k_{c}Y'^{2}+X'^{2}]\:dY'
\end{array}\label{eq:intrinsic-2}
\end{equation}

from which the Jacobian matrix is obtained

\begin{equation}
\begin{array}{c}
\left[\begin{array}{cc}
\frac{\partial X}{\partial X'} & \frac{\partial X}{\partial Y'}\\
\frac{\partial Y}{\partial X'} & \frac{\partial Y}{\partial Y'}
\end{array}\right]=\left[\begin{array}{cc}
f_{x}[1+3k_{c}X'^{2}+Y'^{2}] & \;2f_{x}X'Y^{'}\\
2f_{y}X'Y' & f_{y}[1+3k_{c}Y'^{2}+X'^{2}]
\end{array}\right]\end{array}\label{eq:intrinsic-2-1}
\end{equation}

For any physical point $(x,y,z)$ we need first to calculate the image
coordinates $(X',Y')$ by Eq. \ref{eq:rotation} and \ref{eq:projection},
then use the expressions given above for the Jacobian matrix.

\subsection{Stereoscopic surface mapping}
\label{sub:appendix_mapping}

The first step for stereoscopic view is to map the images of each camera
to the corresponding apparent coordinates on the reference plane.
For that we first determine the physical positions corresponding to
the four corners of the image by the transform Eq. \ref{eq:image2phys}.
We then create a linear grid in the physical space, on which we map
the image values with indices obtained by the transform form physical
to image coordinates. A linear interpolation is used to deal with
non-integer image coordinates. 

Image correlation is then performed between a series of image pairs
from cameras a and b mapped in the apparent physical coordinates on
the reference plane. A regular grid of positions is set on image \emph{a}
and the optimum displacement for image b is obtained. This is performed
in several iterations.

This provides a set of position pairs ($x_{a,},y_{a})$, $(x_{b},y_{b})$
from which the corresponding values of $(x,y,z)$ are obtained from
Eq. \ref{eq:z_stereo}. For simplicity the Jacobian
matrix is calculated on the measurement grid of image \emph{a } for
both cameras and the whole image series, while the
Jacobian matrix for image \emph{b} should be taken at points $(x_{b},y_{b})$. The coefficients  $D_{xb}\sim \mathrm{tan}\beta$  indeed vary slowly 
with position: with  maximum value $z$ equal to 5 cm and a camera at a distance about 5 m, the corresponding change of  $D_{xb}\sim \mathrm{tan}\beta$  is of the order of 1\%. The validity of this approximation has been checked by comparing the results with those obtained by switching the order of the cameras $a$ and $b$. It would be easy to recalculate $D_{xb}$ for each field once the apparent positions on camera b have been determined. 

The PIV procedure involves procedures to remove `false vectors' according
to various criteria: the value of the image correlation, the absence 
of maximum inside the search range and arguments of continuity with
respect to the neighborhood. The error estimate Eq. \ref{eq:error_X}
provides an additional criterion based on the consistency of the stereoscopic
comparison: we eliminate errors bigger than 3 times the rms. For the remaining
data points, the statistics on $\epsilon$ gives a internal evaluation
of the error. Since our cameras are aligned along the $x$ axis we
equivalently use $\epsilon_{Y}$ from Eq. \ref{eq:error_X}. 

After these eliminations of false data, the results on $(x,y,z)$
are eventually interpolated on a regular grid in $(x,y)$, with mesh 1cm, using a
thin plate spline method ~\cite{duchon1977,wahba1990}.

\subsection{3D PIV}

Image correlation is here performed for each camera in a image pair
sliding along the time series with constant time interval. The raw images are used, providing the set of image displacements
$dX_{a},dY_{a},dX_{b},dY_{b}$ in image coordinates. The correlations
are performed from the grid of measurement points on the reference
plane where the $z$ position had been previously determined. This
yields a set of points on images a and b, at which the Jacobian matrix
is calculated as described in Eq. \ref{sub:Jacobian-matrix}. The measured
velocities are however not quite on these points because of the displacement
on the second image. Furthermore some false vectors are eliminated
as described above. Therefore these raw data are interpolated on the
set of image coordinates corresponding to the $(x,y,z)$ fields obtained
at the same time. Then the physical displacements are obtained by
solving the system Eq. \ref{eq:lin_system}, providing a measurement
of the three velocity components. Although the local $z$ displacement
of the surface is used to match the velocity components measurement
by the two cameras, it is not taken into account in the Jacobian matrix as discussed ,in sub-section \ref{sub:appendix_mapping}
which is quite justified as it changes slowly with position. Finally
the error estimate Eq. \ref{eq:error_PIV} is stored as a test of precision.

\bibliographystyle{spmpsci}      % mathematics and physical sciences

%\bibliography{biblio}

\end{document}